\documentclass{article}
\begin{document}
\begin{center}
{\LARGE\bf Hexaquarks in the coupled-channel formalism}

\large

\vskip3ex
S.M. Gerasyuta $ ^{*}$ and E.E. Matskevich $ ^{+}$

\vskip2ex
Department of Theoretical Physics, St. Petersburg State University,
198904,

St. Petersburg, Russia

and

Department of Physics, LTA, 194021, St. Petersburg, Russia

\vskip4ex

{\bf Abstract}
\end{center}
\vskip4ex
\large

The relativistic six-quark equations are found in the framework of the
dispersion relation technique. The approximate solutions of these equations
using the method based on the extraction of leading singularities of the
amplitudes are obtained. The relativistic six-quark amplitudes of
hexaquarks including the quarks of three flavors ($u$, $d$, $s$) are
calculated. The poles of these amplitudes determine the masses of
six-quark systems.

\vskip2ex
\noindent
$ ^{*}$ gerasyuta@SG6488.spb.edu

\noindent
$ ^{+}$ matskev@pobox.spbu.ru
\vskip2ex
\noindent

\vskip2ex
{\bf I. Introduction.}
\vskip2ex

In 1977, Jaffe [1] studied the color-magnetic interaction of the
one-gluon-exchange potential in the multiquark system and found that the
most attractive channel is the flavor singlet with quark content
$u^2d^2s^2$. The same symmetry analysis of the chiral boson exchange
potential leads to the similar result [2].

However, the deuteron channel is not a channel with strong attraction in
any baryon interaction model. If the deuteron had not been found
experimentally, it seems highly unlikely that any model would have been
able to predict it to be a stable dibaryon.

It is shown [3] that there are three types of baryon-baryon bound states.
The states of the first type are called deuteron-like states. If chiral
fields can provide enough attraction between interacting baryons, the
systems would be weakly bound. The states of the second type such as
$\Delta\Delta$, $\Sigma^*\Delta$ are named as $\Delta\Delta$-like states.
Due to highly symmetric character in orbital space, these systems could be
relatively deeply bound, but the strong decay modes of composed baryons
cause the width of the states much broader.

The states of the third type are entitled as $\Omega\Omega$-like states.
Due to the same symmetric character shown in the systems of the second type
and the only weak decay mode of composed baryons, for instance
$\Omega\Omega$, these states are deeply bound states with narrow widths.
The states of latter two types are most interesting new dibaryon states
and should be carefully investigated both theoretically and
experimentally [4 -- 8].

There were number of theoretical predictions by using various models [3],
the quark cluster model [10], the quark-delocation model [11, 12], the
chiral $SU(3)$ quark model [13], the flavor $SU(3)$ skyrmion model [14].
Lomon predicted a deuteron-like dibaryon resonance using R-matrix theory
[15]. By employing the chiral $SU(3)$ quark model Zhang and Yu studied
$\Omega\Omega$ and $\Sigma\Omega$ states [16, 17].

In the series of papers [18 -- 22], a practical treatment of relativistic
three-hadron systems has been developed. The physics of the three-hadron
system is usefully described in terms of the pairwise interactions among
the three particles. The theory is based on the two principles of unitarity
and analyticity, as applied to the two-body subenergy channels. The linear
integral equations in a single variable are obtained for the isobar
amplitudes. Instead of the quadrature methods of obtaining solution the set
of suitable functions are identified and used us a basis for the expansion
of the desired solutions. By this means the coupled integral equations are
solved in terms of simple algebra. In the recent papers [23 -- 25], the
relativistic three-quark equations for the excited baryons are found in
the framework of the dispersion relations technique. We have used the
orbital-spin-flavor functions for the contribution of integral
equations. We searched for the approximate solution of integral
three-quark equations by taking into account two-particle and triangle
singularities, all the weaker ones being neglected. If we considered such
an approximation, which corresponds to taking into account two-body and
triangle singularities, and defined all the smooth functions in the
middle point the physical region of the Dalitz plot, then the problem
was reduced to solving a system of simple algebraic equations.
We calculated the mass spectra of excited baryons using the input
four-fermion interaction with the quantum numbers of gluon [26].

In the present paper the relativistic six-quark equations are found in
the framework of coupled-channel formalism. The dynamical mixing between
the subamplitudes of hexaquark are considered. The six-quark amplitudes
of hexaquarks are calculated. In Sec. II the relativistic six-quark
equations are constructed in the form of the dispersion relation over
the two-body subenergy. The approximate solutions of these equations
using the method based on the extraction of leading singularities of the
amplitude are obtained. Sec. III is devoted to the calculation results for
the hexaquark mass spectra (Tables I and II). In the conclusion, the status
of the considered model is discussed.

\vskip2ex
{\bf II. Six-quark amplitudes of the hexaquarks.}
\vskip2ex

We derive the relativistic six-quark equations in the framework of the
dispersion relation technique. We use only planar diagrams; the other
diagrams due to the rules of $1/N_c$ expansion [27 -- 29] are neglected.
The current generates a six-quark system. The correct equations for the
amplitude are obtained by taking into account all possible subamplitudes.
It corresponds to the division of complete system into subsystems with a
smaller number of particles. Then one should represent a six-particle
amplitude as a sum of 15 subamplitudes:

\begin{equation}
A=\sum\limits_{i<j \atop i=1}^6 A_{ij}\, . \end{equation}

This defines the division of the diagrams into groups according to the
certain pair interaction of particles. The total amplitude can be
represented graphically as a sum of diagrams. We need to consider only
one group of diagrams and the amplitude corresponding to them, for example
$A_{12}$. We shall consider the derivation of the relativistic
generalization of the Faddeev-Yakubovsky approach.

In our case the S-wave hexaquarks are considered. We take into account the
pairwise interaction of all six quarks in the hexaquark.

For instance, we consider the $1^{uu}$-diquarks with spin-parity
$J^P=1^+$ for the hexaquark content ($uuuuuu$) (Fig. 1). The set of
diagrams associated with the amplitude $A_{12}$ can further be broken down
into three groups corresponding to subamplitudes:
$A_1^{1^{uu}}(s,s_{12345},s_{1234},s_{123},s_{12})$,
$A_2^{1^{uu}1^{uu}}(s,s_{12345},s_{1234},s_{12},s_{34})$,
$A_3^{1^{uu}1^{uu}1^{uu}}(s,s_{12345},s_{12},s_{34},s_{56})$.
Here $s_{ik}$ is the two-particle subenergy squared, $s_{ijk}$ corresponds
to the energy squared of particles $i$, $j$, $k$, $s_{ijkl}$ is the energy
squared of particles $i$, $j$, $k$, $l$, $s_{ijklm}$ corresponds to the
energy squared of particles $i$, $j$, $k$, $l$, $m$ and $s$ is the system
total energy squared.

The system of graphical equations is determined by the subamplitudes using
the self-consistent method. The coefficients are determined by the
permutation of quarks [30, 31]. In order to represent the subamplitudes
$A_l$, $l=1-3$ in the form of a dispersion relation, it is necessary to
define the amplitudes of quark-quark interaction. The pair quarks amplitudes
$qq \to qq$ are calculated in the framework of the dispersion $N/D$ method
with the input four-fermion interaction [32 -- 34] with the quantum
numbers of the gluon [26, 35]. The regularization of the dispersion integral
for the $D$-function is carried out with the cutoff parameter $\Lambda$.

The four-quark interaction is considered as an input:

\begin{eqnarray}
 & g_V \left(\bar q \lambda I_f \gamma_{\mu} q \right)^2 +
2\, g^{(s)}_V \left(\bar q \lambda I_f \gamma_{\mu} q \right)
\left(\bar s \lambda \gamma_{\mu} s \right)+
g^{(ss)}_V \left(\bar s \lambda \gamma_{\mu} s \right)^2
 \, . & \end{eqnarray}

\noindent
Here $I_f$ is the unity matrix in the flavor space $(u, d)$, $\lambda$ are
the color Gell-Mann matrices. Dimensional constants of the four-fermion
interaction $g_V$, $g^{(s)}_V$, and $g^{(ss)}_V$ are parameters of the
model.

At $g_V =g^{(s)}_V =g^{(ss)}_V$ the flavor $SU(3)_f$ symmetry occurs.
The strange quark violates the flavor $SU(3)_f$ symmetry. In order to
avoid additional violation parameters we introduce the scale of the
dimensional parameters [35]:

\begin{equation}
g=\frac{m^2}{\pi^2}g_V =\frac{(m+m_s)^2}{4\pi^2}g_V^{(s)} =
\frac{m_s^2}{\pi^2}g_V^{(ss)}
\, .\end{equation}

\begin{equation}
\Lambda=\frac{4\Lambda(ik)}
{(m_i+m_k)^2}. \end{equation}

\noindent
Here $m_i$ and $m_k$ are the quark masses in the intermediate state of
the quark loop. Dimensionless parameters $g$ and $\Lambda$ are supposed
to be constants which are independent of the quark interaction type. The
applicability of Eq. (2) is verified by the success of
De Rujula-Georgi-Glashow quark model [26], where only the short-range
part of Breit potential connected with the gluon exchange is responsible
for the mass splitting in hadron multiplets. We use the results of our
relativistic quark model [35] and write down the pair quark amplitudes in
the form:

\begin{equation}
a_n(s_{ik})=\frac{G^2_n(s_{ik})}
{1-B_n(s_{ik})} \, ,\end{equation}

\begin{equation}
B_n(s_{ik})=\int\limits_{(m_i+m_k)^2}^{\frac{(m_i+m_k)^2\Lambda}{4}}
\hskip2mm \frac{ds'_{ik}}{\pi}\frac{\rho_n(s'_{ik})G^2_n(s'_{ik})}
{s'_{ik}-s_{ik}} \, .\end{equation}

\noindent
Here $G_n(s_{ik})$ are the diquark vertex functions (Table III). The vertex
functions are determined by the contribution of the crossing channels.
The vertex functions satisfy the Fierz relations. All of these vertex
functions are generated from $g_V$, $g^{(s)}_V$ and $g^{(ss)}_V$.
$B_n(s_{ik})$ and $\rho_n (s_{ik})$ are the Chew-Mandelstam functions with
cutoff $\Lambda$ [36] and the phase spaces, respectively:

\begin{eqnarray}
\rho_n (s_{ik},J^{PC})&=&\left(\alpha(n,J^{PC}) \frac{s_{ik}}{(m_i+m_k)^2}
+\beta(n,J^{PC})+\delta(n,J^{PC}) \frac{(m_i-m_k)^2}{s_{ik}}\right)
\nonumber\\
&&\nonumber\\
&\times & \frac{\sqrt{(s_{ik}-(m_i+m_k)^2)(s_{ik}-(m_i-m_k)^2)}}
{s_{ik}}\, .
\end{eqnarray}

The coefficients $\alpha(n,J^{PC})$, $\beta(n,J^{PC})$ and
$\delta(n,J^{PC})$ are given in Table III.

Here $n=1$ coresponds to $qq$-pairs with $J^P=0^+$, $n=2$ corresponds to
$qq$-pairs with $J^P=1^+$.

In the case in question the interacting quarks do not produce a bound
state, therefore the integration in Eqs. (8) -- (10) is carried out from
the threshold $(m_i+m_k)^2$ to the cutoff $\Lambda(ik)$.

We consider the hexaquark state with the strangeness $S=0$, the isospin
$I=3$ (spin-parity $J^P=0^+, 2^+$).

\vskip30pt
\begin{picture}(600,100)
\put(0,45){\line(1,0){18}}
\put(0,47){\line(1,0){17.5}}
\put(0,49){\line(1,0){17}}
\put(0,51){\line(1,0){17}}
\put(0,53){\line(1,0){17.5}}
\put(0,55){\line(1,0){18}}
\put(30,50){\circle{25}}
\put(19,46){\line(1,1){15}}
\put(22,41){\line(1,1){17}}
\put(27.5,38.5){\line(1,1){14}}
\put(31,63){\vector(1,1){20}}
\put(37,60){\vector(1,1){20}}
\put(31,38){\vector(1,-1){20}}
\put(37,40){\vector(1,-1){20}}
\put(50,50){\circle{16}}
\put(58,50){\vector(3,2){22}}
\put(58,50){\vector(3,-2){22}}
\put(75,70){1}
\put(75,23){2}
\put(40,85){4}
\put(60,81){3}
\put(40,10){6}
\put(60,13){5}
\put(42,45){$1^{uu}$}
\put(90,48){$=$}
\put(110,45){\line(1,0){18}}
\put(110,47){\line(1,0){17.5}}
\put(110,49){\line(1,0){17}}
\put(110,51){\line(1,0){17}}
\put(110,53){\line(1,0){17.5}}
\put(110,55){\line(1,0){18}}
\put(135,50){\circle{16}}
\put(128,55){\vector(3,2){22}}
\put(128,55){\vector(2,3){15}}
\put(128,45){\vector(3,-2){22}}
\put(128,45){\vector(2,-3){15}}
\put(143,50){\vector(3,1){22}}
\put(143,50){\vector(3,-1){22}}
\put(167,60){1}
\put(167,33){2}
\put(152,72){3}
\put(140,81){4}
\put(152,21){5}
\put(140,12){6}
\put(127,45){$1^{uu}$}
\put(175,48){$+$}
\put(190,48){8}
\put(203,45){\line(1,0){18}}
\put(203,47){\line(1,0){17.5}}
\put(203,49){\line(1,0){17}}
\put(203,51){\line(1,0){17}}
\put(203,53){\line(1,0){17.5}}
\put(203,55){\line(1,0){18}}
\put(233,50){\circle{25}}
\put(222,46){\line(1,1){15}}
\put(225,41){\line(1,1){17}}
\put(230.5,38.5){\line(1,1){14}}
\put(249,62){\circle{16}}
\put(254.5,68.5){\vector(1,1){15}}
\put(254.5,68.5){\vector(1,-1){18}}
\put(245,50){\vector(1,0){28}}
\put(281,50){\circle{16}}
\put(289,50){\vector(3,1){22}}
\put(289,50){\vector(3,-1){22}}
\put(243,43){\vector(1,-1){15}}
\put(240.5,40.5){\vector(2,-3){12}}
\put(237.5,38){\vector(1,-3){6.5}}
\put(265,61){1}
\put(259,38){2}
\put(310,60){1}
\put(310,33){2}
\put(271,85){3}
\put(261,21){4}
\put(254,12){5}
\put(242,7){6}
\put(273,45){$1^{uu}$}
\put(241,57){$1^{uu}$}
\put(320,48){$+$}
\put(335,48){12}
\put(355,45){\line(1,0){18}}
\put(355,47){\line(1,0){17.5}}
\put(355,49){\line(1,0){17}}
\put(355,51){\line(1,0){17}}
\put(355,53){\line(1,0){17.5}}
\put(355,55){\line(1,0){18}}
\put(385,50){\circle{25}}
\put(374,46){\line(1,1){15}}
\put(377,41){\line(1,1){17}}
\put(382.5,38.5){\line(1,1){14}}
\put(386,63){\vector(1,1){20}}
\put(386,38){\vector(1,-1){20}}
\put(402.5,60){\circle{16}}
\put(402.5,40){\circle{16}}
\put(407,67){\vector(1,1){18}}
\put(407,33){\vector(1,-1){18}}
\put(407,67){\vector(1,-1){17}}
\put(407,33){\vector(1,1){17}}
\put(432,50){\circle{16}}
\put(440,50){\vector(1,1){18}}
\put(440,50){\vector(1,-1){18}}
\put(416,62){1}
\put(416,31){2}
\put(461,65){1}
\put(461,27){2}
\put(429,85){3}
\put(429,10){4}
\put(408,85){5}
\put(408,10){6}
\put(394.5,55){$1^{uu}$}
\put(394.5,35){$1^{uu}$}
\put(424,45){$1^{uu}$}
\end{picture}

\vskip60pt
\begin{picture}(600,60)
\put(0,45){\line(1,0){18}}
\put(0,47){\line(1,0){17.5}}
\put(0,49){\line(1,0){17}}
\put(0,51){\line(1,0){17}}
\put(0,53){\line(1,0){17.5}}
\put(0,55){\line(1,0){18}}
\put(30,50){\circle{25}}
\put(19,46){\line(1,1){15}}
\put(22,41){\line(1,1){17}}
\put(27.5,38.5){\line(1,1){14}}
\put(31,63){\vector(1,1){20}}
\put(31,38){\vector(1,-1){20}}
\put(47.5,60){\circle{16}}
\put(47.5,40){\circle{16}}
\put(55,64){\vector(3,2){18}}
\put(55,36){\vector(3,-2){18}}
\put(55,64){\vector(3,-2){18}}
\put(55,36){\vector(3,2){18}}
\put(78,75){1}
\put(78,53){2}
\put(78,41){3}
\put(78,18){4}
\put(54,80){5}
\put(54,13){6}
\put(39.5,55){$1^{uu}$}
\put(39.5,35){$1^{uu}$}
\put(90,48){$=$}
\put(110,45){\line(1,0){19}}
\put(110,47){\line(1,0){21}}
\put(110,49){\line(1,0){23}}
\put(110,51){\line(1,0){23}}
\put(110,53){\line(1,0){21}}
\put(110,55){\line(1,0){19}}
\put(140,60){\circle{16}}
\put(140,40){\circle{16}}
\put(147.5,64){\vector(3,2){18}}
\put(147.5,36){\vector(3,-2){18}}
\put(147.5,64){\vector(3,-2){18}}
\put(147.5,36){\vector(3,2){18}}
\put(128,55){\vector(1,3){11}}
\put(128,45){\vector(1,-3){11}}
\put(170,75){1}
\put(170,53){2}
\put(170,41){3}
\put(170,18){4}
\put(143,86){5}
\put(143,08){6}
\put(132,55){$1^{uu}$}
\put(132,35){$1^{uu}$}
\put(183,48){$+$}
\put(199,48){4}
\put(212,45){\line(1,0){18}}
\put(212,47){\line(1,0){17.5}}
\put(212,49){\line(1,0){17}}
\put(212,51){\line(1,0){17}}
\put(212,53){\line(1,0){17.5}}
\put(212,55){\line(1,0){18}}
\put(242,50){\circle{25}}
\put(231,46){\line(1,1){15}}
\put(234,41){\line(1,1){17}}
\put(239.5,38.5){\line(1,1){14}}
\put(243,63){\vector(1,1){20}}
\put(243,38){\vector(1,-1){20}}
\put(266,80){5}
\put(266,13){6}
\put(262,50){\circle{16}}
\put(270,50){\vector(3,2){17}}
\put(270,50){\vector(3,-2){17}}
\put(247,61){\vector(1,0){40}}
\put(247,39){\vector(1,0){40}}
\put(295,60){\circle{16}}
\put(295,40){\circle{16}}
\put(303,61){\vector(3,1){20}}
\put(303,61){\vector(3,-1){20}}
\put(303,39){\vector(3,1){20}}
\put(303,39){\vector(3,-1){20}}
\put(328,70){1}
\put(328,53){2}
\put(328,41){3}
\put(328,24){4}
\put(270,66){1}
\put(269,53){2}
\put(268,40){3}
\put(270,28){4}
\put(254,45){$1^{uu}$}
\put(287,55){$1^{uu}$}
\put(287,35){$1^{uu}$}
\put(341,48){$+$}
\put(355,48){8}
\put(368,45){\line(1,0){18}}
\put(368,47){\line(1,0){17.5}}
\put(368,49){\line(1,0){17}}
\put(368,51){\line(1,0){17}}
\put(368,53){\line(1,0){17.5}}
\put(368,55){\line(1,0){18}}
\put(398,50){\circle{25}}
\put(387,46){\line(1,1){15}}
\put(390,41){\line(1,1){17}}
\put(395.5,38.5){\line(1,1){14}}
\put(414,62){\circle{16}}
\put(419.5,68.5){\vector(1,1){15}}
\put(419.5,68.5){\vector(1,-1){18}}
\put(410,50){\vector(1,0){28}}
\put(446,50){\circle{16}}
\put(454,50){\vector(3,1){22}}
\put(454,50){\vector(3,-1){22}}
\put(430,61){1}
\put(427,39){2}
\put(470,62){1}
\put(470,30){2}
\put(436,85){5}
\put(414,37){\circle{16}}
\put(421,32){\vector(3,-1){20}}
\put(421,32){\vector(2,-3){12}}
\put(399,38){\vector(1,-3){8}}
\put(444,22){3}
\put(435,7){4}
\put(409,7){6}
\put(406,57){$1^{uu}$}
\put(438,45){$1^{uu}$}
\put(406,32){$1^{uu}$}
\end{picture}

\vskip60pt
\begin{picture}(600,60)
\put(90,48){$+$}
\put(107,48){4}
\put(125,45){\line(1,0){18}}
\put(125,47){\line(1,0){17.5}}
\put(125,49){\line(1,0){17}}
\put(125,51){\line(1,0){17}}
\put(125,53){\line(1,0){17.5}}
\put(125,55){\line(1,0){18}}
\put(155,50){\circle{25}}
\put(144,46){\line(1,1){15}}
\put(147,41){\line(1,1){17}}
\put(152.5,38.5){\line(1,1){14}}
\put(172.5,60){\circle{16}}
\put(172.5,40){\circle{16}}
\put(177,67){\vector(1,1){18}}
\put(177,33){\vector(1,-1){18}}
\put(177,67){\vector(1,-1){17}}
\put(177,33){\vector(1,1){17}}
\put(202,50){\circle{16}}
\put(210,50){\vector(1,1){18}}
\put(210,50){\vector(1,-1){18}}
\put(155,30){\circle{16}}
\put(155,22){\vector(1,-2){9}}
\put(155,22){\vector(-1,-2){9}}
\put(186,62){1}
\put(186,31){2}
\put(231,65){1}
\put(231,27){2}
\put(199,85){5}
\put(199,10){6}
\put(167,5){3}
\put(137,5){4}
\put(164.5,55){$1^{uu}$}
\put(164.5,35){$1^{uu}$}
\put(194,45){$1^{uu}$}
\put(147,25){$1^{uu}$}
\put(250,48){$+$}
\put(267,48){8}
\put(285,45){\line(1,0){18}}
\put(285,47){\line(1,0){17.5}}
\put(285,49){\line(1,0){17}}
\put(285,51){\line(1,0){17}}
\put(285,53){\line(1,0){17.5}}
\put(285,55){\line(1,0){18}}
\put(315,50){\circle{25}}
\put(304,46){\line(1,1){15}}
\put(307,41){\line(1,1){17}}
\put(312.5,38.5){\line(1,1){14}}
\put(324,68){\circle{16}}
\put(324,32){\circle{16}}
\put(327,53){\vector(3,1){28}}
\put(327,47){\vector(3,-1){28}}
\put(332,70){\vector(3,2){21}}
\put(332,70){\vector(3,-1){23}}
\put(332,30){\vector(3,1){23}}
\put(332,30){\vector(3,-2){21}}
\put(363,60){\circle{16}}
\put(363,40){\circle{16}}
\put(372,60){\vector(3,2){21}}
\put(372,60){\vector(3,-1){23}}
\put(372,40){\vector(3,1){23}}
\put(372,40){\vector(3,-2){21}}
\put(399,71){1}
\put(400,52){2}
\put(400,42){3}
\put(399,19){4}
\put(343,86){5}
\put(343,7){6}
\put(348,68){1}
\put(348,51){2}
\put(348,42){3}
\put(348,25){4}
\put(316,63){$1^{uu}$}
\put(316,27){$1^{uu}$}
\put(355,55){$1^{uu}$}
\put(355,35){$1^{uu}$}
\end{picture}

\vskip60pt
\begin{picture}(600,60)
\put(90,48){$+$}
\put(105,48){16}
\put(125,45){\line(1,0){18}}
\put(125,47){\line(1,0){17.5}}
\put(125,49){\line(1,0){17}}
\put(125,51){\line(1,0){17}}
\put(125,53){\line(1,0){17.5}}
\put(125,55){\line(1,0){18}}
\put(155,50){\circle{25}}
\put(144,46){\line(1,1){15}}
\put(147,41){\line(1,1){17}}
\put(152.5,38.5){\line(1,1){14}}
\put(174,57){\circle{16}}
\put(171,37){\circle{16}}
\put(181,61){\vector(1,1){15}}
\put(156,62.5){\vector(3,1){40}}
\put(204,77){\circle{16}}
\put(181,61){\vector(1,-1){12}}
\put(179,35){\vector(1,1){14}}
\put(201,50){\circle{16}}
\put(209,50){\vector(3,2){18}}
\put(209,50){\vector(3,-2){18}}
\put(212,78){\vector(3,2){18}}
\put(212,78){\vector(3,-2){18}}
\put(179,35){\vector(1,-1){16}}
\put(156,38){\vector(1,-2){11}}
\put(235,87){1}
\put(235,64){2}
\put(231,54){3}
\put(231,32){4}
\put(185,80){1}
\put(190,60){2}
\put(182,46){3}
\put(185,32){4}
\put(198,11){5}
\put(170,9){6}
\put(166,52){$1^{uu}$}
\put(163,32){$1^{uu}$}
\put(196,72){$1^{uu}$}
\put(193,45){$1^{uu}$}
\put(250,48){$+$}
\put(267,48){8}
\put(285,45){\line(1,0){18}}
\put(285,47){\line(1,0){17.5}}
\put(285,49){\line(1,0){17}}
\put(285,51){\line(1,0){17}}
\put(285,53){\line(1,0){17.5}}
\put(285,55){\line(1,0){18}}
\put(315,50){\circle{25}}
\put(304,46){\line(1,1){15}}
\put(307,41){\line(1,1){17}}
\put(312.5,38.5){\line(1,1){14}}
\put(324,68){\circle{16}}
\put(324,32){\circle{16}}
\put(335,50){\circle{16}}
\put(343,50){\vector(1,1){12}}
\put(343,50){\vector(1,-1){12}}
\put(332,70){\vector(3,2){21}}
\put(332,70){\vector(3,-1){23}}
\put(332,30){\vector(3,1){23}}
\put(332,30){\vector(3,-2){21}}
\put(363,60){\circle{16}}
\put(363,40){\circle{16}}
\put(372,60){\vector(3,2){21}}
\put(372,60){\vector(3,-1){23}}
\put(372,40){\vector(3,1){23}}
\put(372,40){\vector(3,-2){21}}
\put(399,71){1}
\put(400,52){2}
\put(400,42){3}
\put(399,19){4}
\put(343,86){5}
\put(343,7){6}
\put(348,68){1}
\put(342,56){2}
\put(342,37){3}
\put(348,25){4}
\put(316,63){$1^{uu}$}
\put(316,27){$1^{uu}$}
\put(327,45){$1^{uu}$}
\put(355,55){$1^{uu}$}
\put(355,35){$1^{uu}$}
\end{picture}

\vskip60pt
\begin{picture}(600,60)
\put(0,45){\line(1,0){18}}
\put(0,47){\line(1,0){17.5}}
\put(0,49){\line(1,0){17}}
\put(0,51){\line(1,0){17}}
\put(0,53){\line(1,0){17.5}}
\put(0,55){\line(1,0){18}}
\put(30,50){\circle{25}}
\put(19,46){\line(1,1){15}}
\put(22,41){\line(1,1){17}}
\put(27.5,38.5){\line(1,1){14}}
\put(40,68){\circle{16}}
\put(40,32){\circle{16}}
\put(51,50){\circle{16}}
\put(48,70){\vector(3,2){19}}
\put(48,70){\vector(3,-1){22}}
\put(59,50){\vector(3,2){14}}
\put(59,50){\vector(3,-2){14}}
\put(48,30){\vector(3,1){22}}
\put(48,30){\vector(3,-2){19}}
\put(70,80){1}
\put(73,63){2}
\put(75,52){3}
\put(75,41){4}
\put(73,30){5}
\put(70,13){6}
\put(32,63){$1^{uu}$}
\put(32,27){$1^{uu}$}
\put(43,45){$1^{uu}$}
\put(90,48){$=$}
\put(110,45){\line(1,0){19}}
\put(110,47){\line(1,0){21}}
\put(110,49){\line(1,0){23}}
\put(110,51){\line(1,0){23}}
\put(110,53){\line(1,0){21}}
\put(110,55){\line(1,0){19}}
\put(133,65){\circle{16}}
\put(142,50){\circle{16}}
\put(133,35){\circle{16}}
\put(141,68){\vector(1,1){13}}
\put(141,68){\vector(3,-1){18}}
\put(150,50){\vector(3,2){16}}
\put(150,50){\vector(3,-2){16}}
\put(141,32){\vector(3,1){18}}
\put(141,32){\vector(1,-1){13}}
\put(143,80){1}
\put(160,65){2}
\put(170,55){3}
\put(170,39){4}
\put(160,28){5}
\put(143,10){6}
\put(125,60){$1^{uu}$}
\put(134,45){$1^{uu}$}
\put(125,30){$1^{uu}$}
\put(183,48){$+$}
\put(196,48){12}
\put(212,45){\line(1,0){18}}
\put(212,47){\line(1,0){17.5}}
\put(212,49){\line(1,0){17}}
\put(212,51){\line(1,0){17}}
\put(212,53){\line(1,0){17.5}}
\put(212,55){\line(1,0){18}}
\put(242,50){\circle{25}}
\put(231,46){\line(1,1){15}}
\put(234,41){\line(1,1){17}}
\put(239.5,38.5){\line(1,1){14}}
\put(242,30){\circle{16}}
\put(242,22){\vector(2,-3){9}}
\put(242,22){\vector(-2,-3){9}}
\put(262,50){\circle{16}}
\put(270,50){\vector(3,2){17}}
\put(270,50){\vector(3,-2){17}}
\put(247,61){\vector(1,0){40}}
\put(247,39){\vector(1,0){40}}
\put(295,60){\circle{16}}
\put(295,40){\circle{16}}
\put(303,61){\vector(3,1){20}}
\put(303,61){\vector(3,-1){20}}
\put(303,39){\vector(3,1){20}}
\put(303,39){\vector(3,-1){20}}
\put(328,70){1}
\put(328,53){2}
\put(328,41){3}
\put(328,24){4}
\put(224,10){5}
\put(255,10){6}
\put(270,66){1}
\put(269,53){2}
\put(268,40){3}
\put(270,28){4}
\put(234,25){$1^{uu}$}
\put(254,45){$1^{uu}$}
\put(287,55){$1^{uu}$}
\put(287,35){$1^{uu}$}
\put(341,48){$+$}
\put(356,48){24}
\put(375,45){\line(1,0){18}}
\put(375,47){\line(1,0){17.5}}
\put(375,49){\line(1,0){17}}
\put(375,51){\line(1,0){17}}
\put(375,53){\line(1,0){17.5}}
\put(375,55){\line(1,0){18}}
\put(405,50){\circle{25}}
\put(394,46){\line(1,1){15}}
\put(397,41){\line(1,1){17}}
\put(402.5,38.5){\line(1,1){14}}
\put(402,62){\vector(2,1){40}}
\put(402,38){\vector(2,-1){40}}
\put(422.5,60){\circle{16}}
\put(422.5,40){\circle{16}}
\put(427,67){\vector(1,1){15}}
\put(427,33){\vector(1,-1){15}}
\put(427,67){\vector(1,-1){17}}
\put(427,33){\vector(1,1){17}}
\put(450,84){\circle{16}}
\put(452,50){\circle{16}}
\put(450,16){\circle{16}}
\put(458,86){\vector(3,2){18}}
\put(458,86){\vector(3,-2){18}}
\put(460,50){\vector(3,2){18}}
\put(460,50){\vector(3,-2){18}}
\put(458,14){\vector(3,2){18}}
\put(458,14){\vector(3,-2){18}}
\put(480,95){1}
\put(480,70){2}
\put(480,58){3}
\put(480,33){4}
\put(480,21){5}
\put(480,-4){6}
\put(425,80){1}
\put(438,68){2}
\put(438,58){3}
\put(438,35){4}
\put(438,24){5}
\put(425,12){6}
\put(414.5,55){$1^{uu}$}
\put(414.5,35){$1^{uu}$}
\put(442,79){$1^{uu}$}
\put(444,45){$1^{uu}$}
\put(442,11){$1^{uu}$}
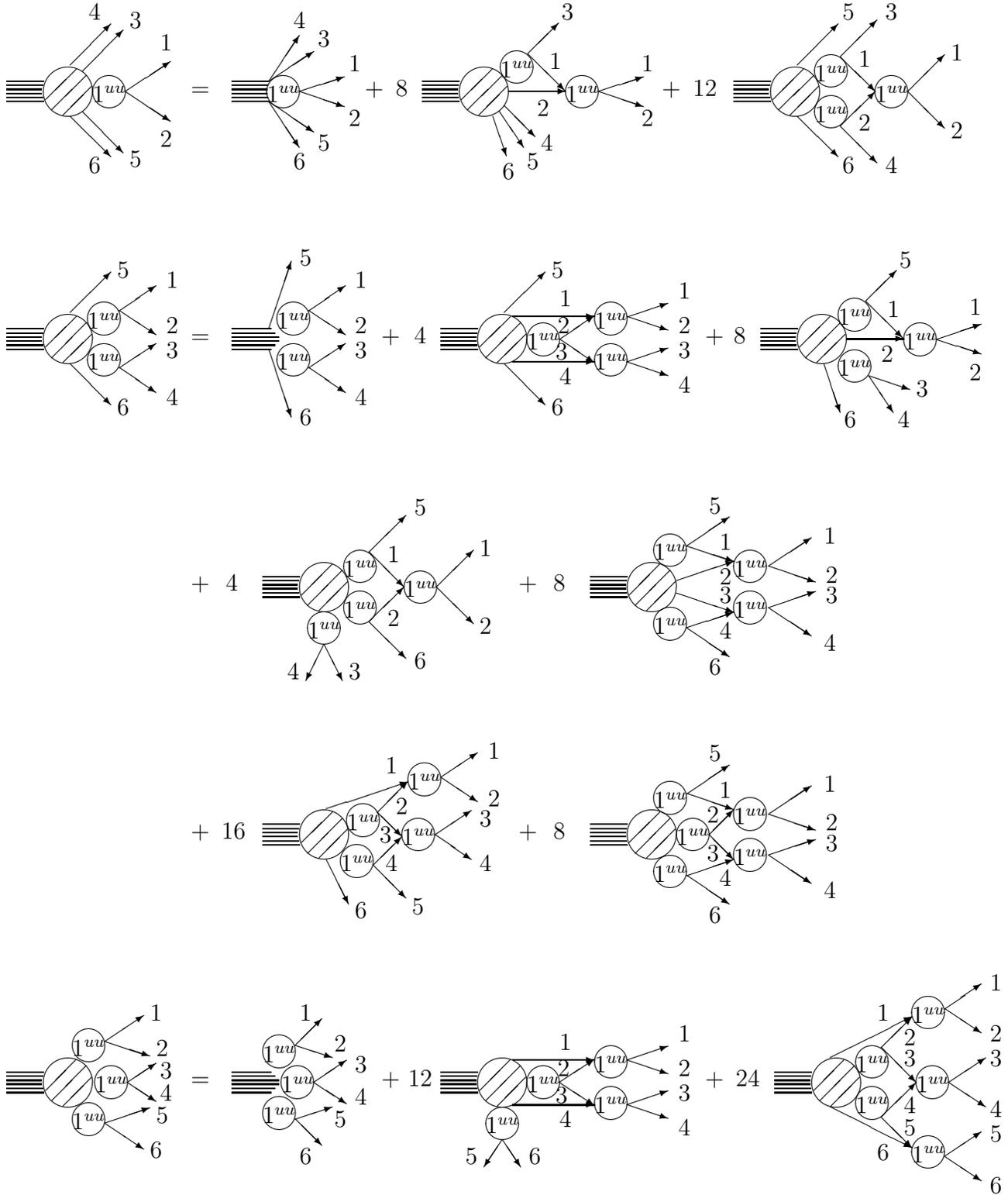
\put(0,-25){Fig. 1. Graphic representation of the equations for the
six-quark subamplitudes $A_l$ $(l=1, 2, 3)$}
\put(0,-45){in the case of the spin-pariry $J^P=0^+, 2^+$ (quark content
($uuuuuu$)).}
\end{picture}

\vskip50pt

The coupled integral equations correspond to Fig. 1 can be described as:

\begin{eqnarray}
A_1^{1^{uu}}(s,s_{12345},s_{1234},s_{123},s_{12})&
=&\frac{\lambda_1 B_{1^{uu}}(s_{12})}{[1-B_{1^{uu}}(s_{12})]}+
8\hat J_1(s_{12},1^{uu}) A_1^{1^{uu}}(s,s_{12345},s_{1234},s_{123},s'_{13})
\nonumber\\
&&\nonumber\\
&+&12\hat J_2(s_{12},1^{uu})
A_2^{1^{uu}1^{uu}}(s,s_{12345},s_{1234},s'_{13},s'_{24}) \, ,
\\
&&\nonumber\\
A_2^{1^{uu}1^{uu}}(s,s_{12345},s_{1234},s_{12},s_{34})&
=&\frac{\lambda_2 B_{1^{uu}}(s_{12})
B_{1^{uu}}(s_{34})}{[1-B_{1^{uu}}(s_{12})][1-B_{1^{uu}}(s_{34})]}
\nonumber\\
&&\nonumber\\
&+&4\hat J_3(s_{12},s_{34},1^{uu},1^{uu})
A_1^{1^{uu}}(s,s_{12345},s_{1234},s'_{123},s'_{23})
\nonumber\\
&&\nonumber\\
&+&8\hat J_4(s_{12},s_{34},1^{uu},1^{uu})
A_1^{1^{uu}}(s,s_{12345},s_{1235},s_{125},s'_{15})
\nonumber\\
&&\nonumber\\
&+&4\hat J_5(s_{12},s_{34},1^{uu},1^{uu})
A_2^{1^{uu}1^{uu}}(s,s_{12356},s_{1256},s'_{15},s'_{26})
\nonumber\\
&&\nonumber\\
&+&8\hat J_6(s_{12},s_{34},1^{uu},1^{uu})
A_2^{1^{uu}1^{uu}}(s,s_{12456},s_{1456},s'_{15},s'_{46})
\nonumber\\
&&\nonumber\\
&+&16\hat J_7(s_{12},s_{34},1^{uu},1^{uu})
A_2^{1^{uu}1^{uu}}(s,s_{12345},s_{2345},s'_{23},s'_{45})
\nonumber\\
&&\nonumber\\
&+&8\hat J_8(s_{12},s_{34},1^{uu},1^{uu})
A_3^{1^{uu}1^{uu}1^{uu}}(s,s_{12345},s'_{15},s'_{23},s'_{46})
 \, ,\\
&&\nonumber\\
A_3^{1^{uu}1^{uu}1^{uu}}(s,s_{12345},s_{12},s_{34},s_{56})&
=&\frac{\lambda_3 B_{1^{uu}}(s_{12})
B_{1^{uu}}(s_{34}) B_{1^{uu}}(s_{56})}{[1- B_{1^{uu}}(s_{12})]
[1- B_{1^{uu}}(s_{34})] [1- B_{1^{uu}}(s_{56})]}
\nonumber \\
&&\nonumber\\
&+&12\hat J_9(s_{12},s_{34},s_{56},1^{uu},1^{uu},1^{uu})
A_1^{1^{uu}}(s,s_{12345},s_{1234},s'_{123},s'_{23})
\nonumber\\
&&\nonumber\\
&+&24\hat J_{10}(s_{12},s_{34},s_{56},1^{uu},1^{uu},1^{uu})
A_2^{1^{uu}1^{uu}}(s,s_{12345},s_{2345},s'_{23},s'_{45})\, ,\nonumber \\
&&
\end{eqnarray}

\noindent
where

\begin{eqnarray}
\hat J_1(s_{12},i)&=&\frac{G_i(s_{12})}{[1- B_i(s_{12})]}
\int\limits_{(m_1+m_2)^2}^{\frac{(m_1+m_2)^2\Lambda_i}{4}}
\frac{ds'_{12}}{\pi}\frac{G_i(s'_{12})\rho_i(s'_{12})}
{s'_{12}-s_{12}} \int\limits_{-1}^{+1} \frac{dz_1(1)}{2} \, ,\\
&&\nonumber\\
\hat J_2(s_{12},i)&=&\frac{G_i(s_{12})}{[1- B_i(s_{12})]}
\int\limits_{(m_1+m_2)^2}^{\frac{(m_1+m_2)^2\Lambda_i}{4}}
\frac{ds'_{12}}{\pi}\frac{G_i(s'_{12})\rho_i(s'_{12})}
{s'_{12}-s_{12}}
\frac{1}{2\pi}\int\limits_{-1}^{+1}\frac{dz_1(2)}{2}
\int\limits_{-1}^{+1} \frac{dz_2(2)}{2}
\nonumber\\
&&\nonumber\\
&\times&
\int\limits_{z_3(2)^-}^{z_3(2)^+} dz_3(2)
\frac{1}{\sqrt{1-z_1^2(2)-z_2^2(2)-z_3^2(2)+2z_1(2) z_2(2) z_3(2)}} \, , \\
&&\nonumber\\
\hat J_3(s_{12},s_{34},i,j)&=&\frac{G_i(s_{12})G_j(s_{34})}
{[1- B_i(s_{12})][1- B_j(s_{34})]}
\int\limits_{(m_1+m_2)^2}^{\frac{(m_1+m_2)^2\Lambda_i}{4}}
\frac{ds'_{12}}{\pi}\frac{G_i(s'_{12})\rho_i(s'_{12})}
{s'_{12}-s_{12}}\nonumber\\
&&\nonumber\\
&\times&\int\limits_{(m_3+m_4)^2}^{\frac{(m_3+m_4)^2\Lambda_j}{4}}
\frac{ds'_{34}}{\pi}\frac{G_j(s'_{34})\rho_j(s'_{34})}
{s'_{34}-s_{34}}
\int\limits_{-1}^{+1} \frac{dz_1(3)}{2} \int\limits_{-1}^{+1}
\frac{dz_2(3)}{2} \, , \\
&&\nonumber\\
\hat J_4(s_{12},s_{34},i,j)&=&\frac{B_j(s_{34})}{[1- B_j(s_{34})]}
\hat J_1(s_{12},i) \, , \\
&&\nonumber\\
\hat J_5(s_{12},s_{34},i,j)&=&\frac{B_j(s_{34})}{[1- B_j(s_{34})]}
\hat J_2(s_{12},i) \, , \\
&&\nonumber\\
\hat J_6(s_{12},s_{34},i,j)&=&\hat J_1(s_{12},i) \cdot \hat J_1(s_{34},j)
 \, , \\
&&\nonumber\\
\hat J_7(s_{12},s_{34},i,j)&=&\frac{G_i(s_{12})G_j(s_{34})}
{[1- B_i(s_{12})][1- B_j(s_{34})]}
\int\limits_{(m_1+m_2)^2}^{\frac{(m_1+m_2)^2\Lambda_i}{4}}
\frac{ds'_{12}}{\pi}\frac{G_i(s'_{12})\rho_i(s'_{12})}
{s'_{12}-s_{12}}\nonumber\\
&&\nonumber\\
&\times&\int\limits_{(m_3+m_4)^2}^{\frac{(m_3+m_4)^2\Lambda_j}{4}}
\frac{ds'_{34}}{\pi}\frac{G_j(s'_{34})\rho_j(s'_{34})}
{s'_{34}-s_{34}}
\frac{1}{2\pi}\int\limits_{-1}^{+1}\frac{dz_1(7)}{2}
\int\limits_{-1}^{+1} \frac{dz_2(7)}{2}
\int\limits_{-1}^{+1} \frac{dz_3(7)}{2}
\nonumber\\
&&\nonumber\\
&\times&
\int\limits_{z_4(7)^-}^{z_4(7)^+} dz_4(7)
\frac{1}{\sqrt{1-z_1^2(7)-z_3^2(7)-z_4^2(7)+2z_1(7) z_3(7) z_4(7)}}
 \, , \\
&&\nonumber\\
\hat J_8(s_{12},s_{34},i,j)&=&\frac{G_i(s_{12})G_j(s_{34})}
{[1- B_i(s_{12})][1- B_j(s_{34})]}
\int\limits_{(m_1+m_2)^2}^{\frac{(m_1+m_2)^2\Lambda_i}{4}}
\frac{ds'_{12}}{\pi}\frac{G_i(s'_{12})\rho_i(s'_{12})}
{s'_{12}-s_{12}}\nonumber\\
&&\nonumber\\
&\times&\int\limits_{(m_3+m_4)^2}^{\frac{(m_3+m_4)^2\Lambda_j}{4}}
\frac{ds'_{34}}{\pi}\frac{G_j(s'_{34})\rho_j(s'_{34})}
{s'_{34}-s_{34}}\nonumber\\
&&\nonumber\\
&\times&\frac{1}{(2\pi)^2}\int\limits_{-1}^{+1}\frac{dz_1(8)}{2}
\int\limits_{-1}^{+1} \frac{dz_2(8)}{2}
\int\limits_{-1}^{+1} \frac{dz_3(8)}{2}
\int\limits_{z_4(8)^-}^{z_4(8)^+} dz_4(8)
\int\limits_{-1}^{+1} \frac{dz_5(8)}{2}
\int\limits_{z_6(8)^-}^{z_6(8)^+} dz_6(8)
\nonumber\\
&&\nonumber\\
&\times&
\frac{1}{\sqrt{1-z_1^2(8)-z_3^2(8)-z_4^2(8)+2z_1(8) z_3(8) z_4(8)}}
\nonumber\\
&&\nonumber\\
&\times&
\frac{1}{\sqrt{1-z_2^2(8)-z_5^2(8)-z_6^2(8)+2z_2(8) z_5(8) z_6(8)}}
 \, , \\
&&\nonumber\\
\hat J_9(s_{12},s_{34},s_{56},i,j,k)&=&\frac{B_k(s_{56})}{[1- B_k(s_{56})]}
\hat J_3(s_{12},s_{34},i,j) \, , \\
&&\nonumber\\
\hat J_{10}(s_{12},s_{34},s_{56},i,j,k)&=
&\frac{G_i(s_{12})G_j(s_{34})G_k(s_{56})}
{[1- B_i(s_{12})][1- B_j(s_{34})][1- B_k(s_{56})]}
\int\limits_{(m_1+m_2)^2}^{\frac{(m_1+m_2)^2\Lambda_i}{4}}
\frac{ds'_{12}}{\pi}\frac{G_i(s'_{12})\rho_i(s'_{12})}{s'_{12}-s_{12}}
 \nonumber\\
&&\nonumber\\
&\times&
\int\limits_{(m_3+m_4)^2}^{\frac{(m_3+m_4)^2\Lambda_j}{4}}
\frac{ds'_{34}}{\pi}\frac{G_j(s'_{34})\rho_j(s'_{34})}
{s'_{34}-s_{34}}\int\limits_{(m_5+m_6)^2}^{\frac{(m_5+m_6)^2\Lambda_k}{4}}
\frac{ds'_{56}}{\pi}\frac{G_k(s'_{56})\rho_k(s'_{56})}{s'_{56}-s_{56}}
\nonumber\\
&&\nonumber\\
&\times&
\frac{1}{2\pi}\int\limits_{-1}^{+1}\frac{dz_1(10)}{2}
\int\limits_{-1}^{+1} \frac{dz_2(10)}{2}
\int\limits_{-1}^{+1} \frac{dz_3(10)}{2}
\int\limits_{-1}^{+1} \frac{dz_4(10)}{2}
\int\limits_{z_5(1-)^-}^{z_5(10)^+} dz_5(10)
\nonumber\\
&&\nonumber\\
&\times&
\frac{1}{\sqrt{1-z_1^2(10)-z_4^2(10)-z_5^2(10)+2z_1(10) z_4(10) z_5(10)}}
 \, .
\end{eqnarray}

We should discuss the coefficients multiplying of the diagrams in the
equations of Fig. 1. For example, we consider the first subamplitude
$A_1(s,s_{12345},s_{1234},s_{123},s_{12})$. In the Eq. (8) (Fig. 1) the
first coefficient is equal to 8, that the number $8=2$ (permutation
particles 1 and 2) $\times 4$ (we can use third, 4-th, 5-th, 6-th
particles); the second coefficient equal to 12 that the number
$12=4$ (used third, 4-th, 5-th, 6-th particles) $\times 3$ (in this case
we can consider 4-th, 5-th, 6-th particles). The similar approach allows
us to take into account the coefficients in the Eqs. (9) and (10).

Let us extract two- and three-particle singularities in the amplitudes
$A_1^{1^{uu}}(s,s_{12345},s_{1234},s_{123},s_{12})$,
$A_2^{1^{uu}1^{uu}}(s,s_{12345},s_{1234},s_{12},s_{34})$,
$A_3^{1^{uu}1^{uu}1^{uu}}(s,s_{12345},s_{12},s_{34},s_{56})$:

\begin{eqnarray}
A_1^{1^{uu}}(s,s_{12345},s_{1234},s_{123},s_{12})&=
&\frac{\alpha_1^{1^{uu}} (s,s_{12345},s_{1234},s_{123},s_{12})
B_{1^{uu}}(s_{12})}{[1-B_{1^{uu}}(s_{12})]} \, ,\\
&&\nonumber\\
A_2^{1^{uu}1^{uu}}(s,s_{12345},s_{1234},s_{12},s_{34})&=
&\frac{\alpha_2^{1^{uu}1^{uu}} (s,s_{12345},s_{1234},s_{12},s_{34})
B_{1^{uu}}(s_{12})B_{1^{uu}}(s_{34})}{[1-B_{1^{uu}}(s_{12})]
[1-B_{1^{uu}}(s_{34})]} \, , \\
&&\nonumber\\
A_3^{1^{uu}1^{uu}1^{uu}}(s,s_{12345},s_{12},s_{34},s_{56})&=
&\frac{\alpha_3^{1^{uu}1^{uu}1^{uu}} (s,s_{12345},s_{12},s_{34},s_{56})
B_{1^{uu}}(s_{12})B_{1^{uu}}(s_{34}) B_{1^{uu}}(s_{56})}
{[1- B_{1^{uu}}(s_{12})] [1- B_{1^{uu}}(s_{34})][1- B_{1^{uu}}(s_{56})]}
 \, . \nonumber\\
&&
\end{eqnarray}

We do not extract four-particles singularities, because they are weaker
than two- and three-particle singularities.

We used the classification of singularities, which was proposed in
paper [37]. The construction of the approximate solution of Eqs.
(21) -- (23) is based on the extraction of the leading singularities
of the amplitudes. The main singularities in $s_{ik}=(m_i+m_k)^2$
are from pair rescattering of the particles $i$ and $k$. First of all there
are threshold square-root singularities. Also possible are pole
singularities which correspond to the bound states. The diagrams of Fig. 1
apart two-particle singularities have triangular singularities and the
singularities defining the interactions of four, five and six particles.
Such classification allows us to search the corresponding solution of Eqs.
(8) -- (10) by taking into account some definite number of leading
singularities and neglecting all the weaker ones. We consider the
approximation which defines two-particle, triangle and four-,  five- and
six-particle singularities. The contribution of two-particle and triangle
singularities are more important, but we must take into account also the
other singularities.

The functions $\alpha_l$, $l=1-3$ are the smooth functions of $s_{ik}$,
$s_{ijk}$, $s_{ijkl}$ $s_{ijklm}$ as compared with the singular part of the
amplitudes, hence they can be expanded in a series should be employed
further. Using this classification, one defines the reduced amplitudes
$\alpha_1$, $\alpha_2$, $\alpha_3$ as well as the $B$-functions
in the middle point of physical region of Dalitz-plot at the point $s_0$:

\begin{eqnarray}
s_0=\frac{s+4\sum\limits_{i=1}^{6} m_i^2}
{\sum\limits_{i,k=1 \atop i<k}^{6} m_{ik}^2}
\, ,
\end{eqnarray}

\begin{eqnarray}
s_{123}=s_0 \sum\limits_{i,k=1 \atop i<k}^{3} m_{ik}^2
-\sum\limits_{i=1}^{3} m_i^2
\, ,
\end{eqnarray}

\begin{eqnarray}
s_{1234}=s_0 \sum\limits_{i,k=1 \atop i<k}^{4} m_{ik}^2
-2\sum\limits_{i=1}^{4} m_i^2
\, .
\end{eqnarray}

Such choice of point $s_0$ allows us to replace integral equations
(8) -- (10) (Fig. 1) by the algebraic equations (27) -- (29), respectively:

\begin{eqnarray}
\alpha_1^{1^{uu}} &=&\lambda+8 I_1(1^{uu}1^{uu}) \alpha_1^{1^{uu}}
+12 I_2(1^{uu}1^{uu}1^{uu}) \alpha_2^{1^{uu}1^{uu}}
\, , \\
&&\nonumber\\
\alpha_2^{1^{uu}1^{uu}} &=&\lambda+4 I_3(1^{uu}1^{uu}1^{uu})
\alpha_1^{1^{uu}}+8 I_4(1^{uu}1^{uu}1^{uu}) \alpha_1^{1^{uu}}
+4 I_5(1^{uu}1^{uu}1^{uu}1^{uu}) \alpha_2^{1^{uu}1^{uu}}\nonumber\\
&&\nonumber\\
&+&8 I_6(1^{uu}1^{uu}1^{uu}1^{uu}) \alpha_2^{1^{uu}1^{uu}}
+16 I_7(1^{uu}1^{uu}1^{uu}1^{uu}) \alpha_2^{1^{uu}1^{uu}}
+8 I_8(1^{uu}1^{uu}1^{uu}1^{uu}1^{uu}) \alpha_3^{1^{uu}1^{uu}1^{uu}}
\, , \nonumber\\
&& \\
\alpha_3^{1^{uu}1^{uu}1^{uu}} &=&\lambda+12 I_9(1^{uu}1^{uu}1^{uu}1^{uu})
\alpha_1^{1^{uu}}+24 I_{10}(1^{uu}1^{uu}1^{uu}1^{uu}1^{uu})
\alpha_2^{1^{uu}1^{uu}} \, ,
\end{eqnarray}

\noindent
where $\lambda_i$ are the current constants. We used the functions
$I_1$, $I_2$, $I_3$, $I_4$, $I_5$, $I_6$, $I_7$, $I_8$, $I_9$, $I_{10}$:

\begin{eqnarray}
I_1(ij)&=&\frac{B_j(s_0^{13})}{B_i(s_0^{12})}
\int\limits_{(m_1+m_2)^2}^{\frac{(m_1+m_2)^2\Lambda_i}{4}}
\frac{ds'_{12}}{\pi}\frac{G_i^2(s_0^{12})\rho_i(s'_{12})}
{s'_{12}-s_0^{12}} \int\limits_{-1}^{+1} \frac{dz_1(1)}{2}
\frac{1}{1-B_j (s'_{13})}\, , \\
&&\nonumber\\
I_2(ijk)&=&\frac{B_j(s_0^{13}) B_k(s_0^{24})}{B_i(s_0^{12})}
\int\limits_{(m_1+m_2)^2}^{\frac{(m_1+m_2)^2\Lambda_i}{4}}
\frac{ds'_{12}}{\pi}\frac{G_i^2(s_0^{12})\rho_i(s'_{12})}
{s'_{12}-s_0^{12}}
\frac{1}{2\pi}\int\limits_{-1}^{+1}\frac{dz_1(2)}{2}
\int\limits_{-1}^{+1} \frac{dz_2(2)}{2}\nonumber\\
&&\nonumber\\
&\times&
\int\limits_{z_3(2)^-}^{z_3(2)^+} dz_3(2)
\frac{1}{\sqrt{1-z_1^2(2)-z_2^2(2)-z_3^2(2)+2z_1(2) z_2(2) z_3(2)}}
\nonumber\\
&&\nonumber\\
&\times& \frac{1}{1-B_j (s'_{13})} \frac{1}{1-B_k (s'_{24})}
 \, , \\
&&\nonumber\\
I_3(ijk)&=&\frac{B_k(s_0^{23})}{B_i(s_0^{12}) B_j(s_0^{34})}
\int\limits_{(m_1+m_2)^2}^{\frac{(m_1+m_2)^2\Lambda_i}{4}}
\frac{ds'_{12}}{\pi}\frac{G_i^2(s_0^{12})\rho_i(s'_{12})}
{s'_{12}-s_0^{12}}\nonumber\\
&&\nonumber\\
&\times&\int\limits_{(m_3+m_4)^2}^{\frac{(m_3+m_4)^2\Lambda_j}{4}}
\frac{ds'_{34}}{\pi}\frac{G_j^2(s_0^{34})\rho_j(s'_{34})}
{s'_{34}-s_0^{34}}
\int\limits_{-1}^{+1} \frac{dz_1(3)}{2} \int\limits_{-1}^{+1}
\frac{dz_2(3)}{2} \frac{1}{1-B_k (s'_{23})} \, , \\
&&\nonumber\\
I_4(ijk)&=&I_1(ik) \, , \\
&&\nonumber\\
I_5(ijkl)&=&I_2(ikl) \, , \\
&&\nonumber\\
I_6(ijkl)&=&I_1(ik) \cdot I_1(jl)
 \, , \\
&&\nonumber\\
I_7(ijkl)&=&\frac{B_k(s_0^{23})B_l(s_0^{45})}{B_i(s_0^{12}) B_j(s_0^{34})}
\int\limits_{(m_1+m_2)^2}^{\frac{(m_1+m_2)^2\Lambda_i}{4}}
\frac{ds'_{12}}{\pi}\frac{G_i^2(s_0^{12})\rho_i(s'_{12})}
{s'_{12}-s_0^{12}}\nonumber\\
&&\nonumber\\
&\times&\int\limits_{(m_3+m_4)^2}^{\frac{(m_3+m_4)^2\Lambda_j}{4}}
\frac{ds'_{34}}{\pi}\frac{G_j^2(s_0^{34})\rho_j(s'_{34})}
{s'_{34}-s_{34}}
\frac{1}{2\pi}\int\limits_{-1}^{+1}\frac{dz_1(7)}{2}
\int\limits_{-1}^{+1} \frac{dz_2(7)}{2}
\int\limits_{-1}^{+1} \frac{dz_3(7)}{2}
\nonumber\\
&&\nonumber\\
&\times&
\int\limits_{z_4(7)^-}^{z_4(7)^+} dz_4(7)
\frac{1}{\sqrt{1-z_1^2(7)-z_3^2(7)-z_4^2(7)+2z_1(7) z_3(7) z_4(7)}}
\nonumber\\
&&\nonumber\\
&\times& \frac{1}{1-B_k (s'_{23})} \frac{1}{1-B_l (s'_{45})}
 \, , \\
&&\nonumber\\
I_8(ijklm)&=&\frac{B_k(s_0^{15})B_l(s_0^{23})B_m(s_0^{46})}
{B_i(s_0^{12}) B_j(s_0^{34})}
\int\limits_{(m_1+m_2)^2}^{\frac{(m_1+m_2)^2\Lambda_i}{4}}
\frac{ds'_{12}}{\pi}\frac{G_i^2(s_0^{12})\rho_i(s'_{12})}
{s'_{12}-s_0^{12}}\nonumber\\
&&\nonumber\\
&\times&\int\limits_{(m_3+m_4)^2}^{\frac{(m_3+m_4)^2\Lambda_j}{4}}
\frac{ds'_{34}}{\pi}\frac{G_j^2(s_0^{34})\rho_j(s'_{34})}
{s'_{34}-s_0^{34}}\nonumber\\
&&\nonumber\\
&\times&\frac{1}{(2\pi)^2}\int\limits_{-1}^{+1}\frac{dz_1(8)}{2}
\int\limits_{-1}^{+1} \frac{dz_2(8)}{2}
\int\limits_{-1}^{+1} \frac{dz_3(8)}{2}
\int\limits_{z_4(8)^-}^{z_4(8)^+} dz_4(8)
\int\limits_{-1}^{+1} \frac{dz_5(8)}{2}
\int\limits_{z_6(8)^-}^{z_6(8)^+} dz_6(8)
\nonumber\\
&&\nonumber\\
&\times&
\frac{1}{\sqrt{1-z_1^2(8)-z_3^2(8)-z_4^2(8)+2z_1(8) z_3(8) z_4(8)}}
\nonumber\\
&&\nonumber\\
&\times&
\frac{1}{\sqrt{1-z_2^2(8)-z_5^2(8)-z_6^2(8)+2z_2(8) z_5(8) z_6(8)}}
\nonumber\\
&&\nonumber\\
&\times& \frac{1}{1-B_k (s'_{15})} \frac{1}{1-B_l (s'_{23})}
\frac{1}{1-B_m (s'_{46})}
 \, , \\
&&\nonumber\\
I_9(ijkl)&=&I_3(ijl) \, , \\
&&\nonumber\\
I_{10}(ijklm)&=
&\frac{B_l(s_0^{23})B_m(s_0^{45})}
{B_i(s_0^{12}) B_j(s_0^{34}) B_k(s_0^{56})}
\int\limits_{(m_1+m_2)^2}^{\frac{(m_1+m_2)^2\Lambda_i}{4}}
\frac{ds'_{12}}{\pi}\frac{G_i^2(s_0^{12})\rho_i(s'_{12})}{s'_{12}-s_0^{12}}
\nonumber\\
&&\nonumber\\
&\times&
\int\limits_{(m_3+m_4)^2}^{\frac{(m_3+m_4)^2\Lambda_j}{4}}
\frac{ds'_{34}}{\pi}\frac{G_j^2(s_0^{34})\rho_j(s'_{34})}
{s'_{34}-s_0^{34}}
\int\limits_{(m_5+m_6)^2}^{\frac{(m_5+m_6)^2\Lambda_k}{4}}
\frac{ds'_{56}}{\pi}\frac{G_k^2(s_0^{56})\rho_k(s'_{56})}{s'_{56}-s_0^{56}}
\nonumber\\
&&\nonumber\\
&\times&
\frac{1}{2\pi}\int\limits_{-1}^{+1}\frac{dz_1(10)}{2}
\int\limits_{-1}^{+1} \frac{dz_2(10)}{2}
\int\limits_{-1}^{+1} \frac{dz_3(10)}{2}
\int\limits_{-1}^{+1} \frac{dz_4(10)}{2}
\int\limits_{z_5(1-)^-}^{z_5(10)^+} dz_5(10)
\nonumber\\
&&\nonumber\\
&\times&
\frac{1}{\sqrt{1-z_1^2(10)-z_4^2(10)-z_5^2(10)+2z_1(10) z_4(10) z_5(10)}}
\nonumber\\
&&\nonumber\\
&\times& \frac{1}{1-B_l (s'_{23})} \frac{1}{1-B_m (s'_{45})}
 \, ,
\end{eqnarray}

\noindent
where $i$, $j$, $k$, $l$, $m$ correspond to the diquarks with the
spin-parity $J^P=0^+, 1^+$.

In the equation (30) $z_1(1)$ is the cosine of the angle between the
relative momentum of particles 1 and 2 in the intermediate state and the
momentum of the particle 3 in the final state taken in the c.m. of particles
1 and 2. We can go from the integration of the cosine of the angle
$dz_1(1)$ to the integration over the subenergy $ds'_{13}$.

In Eq. (31) $z_1(2)$ is the cosine of the angle between the
relative momentum of particles 1 and 2 in the intermediate state and the
momentum of the particle 3 in the final state taken in the c.m. of particles
1 and 2, $z_2(2)$ is the cosine of the angle between the momenta of
particles 3 and 4 in the final state of c.m. of particles 1 and 2,
$z_3(2)$ is cosine of the angle between the relative momentum of particles
1 and 2 in the intermediate state and the momentum of the particle 4 in the
final state of c.m. of particles 1 and 2. Then we pass from
$dz_1(2)dz_2(2)dz_3(2)$ to $ds'_{13}ds'_{34}ds'_{24}$.

In Eq. (32) $z_1(3)$ is the cosine of the angle between the relative
momentum of particles 1, 2 in the intermediate state and the relative
momentum of particles 3, 4 in the intermediate state in c.m. of particles
3 and 4; $z_2(3)$ is the cosine of the angle between momentum of particle 3
in the intermediate state and relative momentum of particles 1, 2 in the
intermediate state in c.m. 1 and 2. We pass from $dz_1(3)dz_2(3)$ to
$ds'_{123}ds'_{23}$. The similar method are used for the functions
(33) -- (35), (38).

In Eq. (36) $z_1(7)$ is cosine of the angle between relative momentum
of the particles 1, 2 in the intermediate state and the relative momentum of
particles 3, 4 in the intermediate state in c.m. of particles 3 and 4;
$z_2(7)$ is the cosine of the angle between the momentum of particle 3
in the intermediate state and relative momentum of particles 1, 2 in the
intermediate state in c.m. of particles 1 and 2; $z_3(7)$ is cosine of the
angle between momentum of particle 5 in the final state and relative
momentum of particles 1, 2 in the intermediate state in c.m. of particles
3 and 4; $z_4(7)$ is cosine of the angle between momentum of particle 5 in
the final state and the relative momentum of particles 3, 4 in the
intermediate state in c.m. of particles 3 and 4. Then we translated the
$dz_1(7)dz_2(7)dz_3(7)dz_4(7)$ to $ds'_{123}ds'_{23}ds'_{125}ds'_{45}$.

In Eq. (37) $z_1(8)$ is the cosine of the angle between momentum of particle
5 in the final state and the relative momentum of particles 1, 2 in the
intermediate state in c.m. of particles 1 and 2; $z_2(8)$ is the cosine
of the angle between the relative momentum of particles 1, 2 in the
intermediate state and the relative momentum of particles 3, 4 in the
intermediate state in c.m. of particles 3 and 4; $z_3(8)$ is the cosine of
the angle between momentum of particle 3 in the intermediate state and the
momentum of particle 5 in the final state in c.m. of particles 1 and 2;
$z_4(8)$ is the cosine of the angle between the momentum of particle 3 in
the intermediate state and the relative momentum of particles 1, 2 in the
intermediate state in c.m. of particles 1 and 2; $z_5(8)$ is the cosine
of angle between momentum of particle 6 in the final state and the relative
momentum of particles 1, 2 in the intermediate state in c.m. of particles
3 and 4; $z_6(8)$ is the cosine of the angle between momentum of particle
6 in the final state and the relative momentum of particles 3, 4 in the
intermediate state in c.m. of particles 3 and 4. We pass from
$dz_1(8)dz_2(8)dz_3(8)dz_4(8)dz_5(8)dz_6(8)$ to
$ds'_{15}ds'_{123}ds'_{35}ds'_{23}ds'_{126}ds'_{46}$.

In Eq. (39) $z_1(10)$ is the cosine of angle between relative momentum of
particles 1, 2 in the intermediate state and the relative momentum of
particles 3, 4 in the intermediate state in c.m. of particles 3 and 4;
$z_2(10)$ is the cosine of angle between the relative momentum of particles
1, 2 in the intermediate state and momentum of particle 3 in the final state
in c.m. of particles 1 and 2; $z_3(10)$ is the cosine of the angle between
the relative momentum of the particles 3, 4 in the intermediate state and
the relative momentum of particles 5, 6 in the intermediate state in c.m.
of particles 5 and 6; $z_4(10)$ is the cosine of angle between relative
momentum of particles 1, 2 in the intermediate state and the momentum of
particle 5 in the final state in c.m. of particles 3 and 4; $z_5(10)$ is
the cosine of the angle between the relative momentum of the particles 3, 4
in the intermediate state and the momentum of particle 5 in the final state
in c.m. of particles 3 and 4. We pass from
$dz_1(10)dz_2(10)dz_3(10)dz_4(10)dz_5(10)$ to
$ds'_{123}ds'_{23}ds'_{345}ds'_{125}ds'_{45}$.

The other choices of point $s_0$ do not change essentially the contributions
of $\alpha_l$, $l=1-3$, therefore we omit the indices $s_0^{ik}$. Since the
vertex functions depend only slightly on energy, it is possible to treat
them as constants in our approximation.

The solutions of the system of equations are considered as:

\begin{equation}
\alpha_i(s)=F_i(s,\lambda_i)/D(s) \, ,\end{equation}

\noindent
where zeros of $D(s)$ determinants define the masses of bound states of
dibaryons.

As example, we consider the equations for the quark content $uuuuuu$ with
the strangeness $S=0$, the isospin $I=3$ and the spin-parity
$J^P=0^+, 2^+$ (Fig. 1). The similar equations have been
calculated for the strangeness $S=0, -1, -2, -3, -4, -5, -6$,
the isospin $I=0$, $\frac{1}{2}$, $1$, $\frac{3}{2}$, $2$, $\frac{5}{2}$,
$3$ and the spin-parity $J^P=0^+, 1^+, 2^+$. We take into account the
$u$, $d$, $s$ quarks.

In Appendix I the $NN_{SIJ=001}$, $\Delta\Delta_{SIJ=001}$,
$\Omega\Omega_{SIJ=-600}$, $\Lambda\Lambda_{SIJ=-200}$,
$N\Omega_{SIJ=-3 \frac{1}{2} 2}$ dibaryons are given.

\vskip2ex
{\bf III. Calculation results.}
\vskip2ex

The poles of the reduced amplitudes $\alpha_i$ ($i=1-3$) correspond to the
bound state and determine the mass of the hexaquark with the quark content
($uuuuuu$), with the isospin $I=3$ and the spin-parity $J^P=0^+, 2^+$.
The quark masses of model $m_{u,d}=410\, MeV$ and $m_s=557\, MeV$ coincide
with the ordinary baryon ones in our model [38].

The model in question has only three parameters: the cutoff parameter
$\Lambda=11$ (similar to the model [41]) and the gluon coupling constants
$g_0$ and $g_1$. These parameters are determined by the $\Lambda\Lambda$ and
the di-$\Omega$ masses. We have considered the two type of calculations. In
the first case we use the gluon coupling constants $g_1=0.292$
(diquark $1^+$) and $g_0=0.653$ (diquark $0^+$). Which are fitted by the
$\Lambda\Lambda$ state with the $M=2173\, MeV$ and the di-$\Omega$ with the
$M=3232\, MeV$, respectively. In the second case the gluon coupling
constants $g_1=0.325$ and $g_0=0.647$ are determined by the masses of
$\Lambda\Lambda$ state with the $M=2171\, MeV$ and the di-$\Omega$ state
$M=3093\, MeV$. The experimental data of these masses are absent, therefore
we use the paper [12]. In our model the correlation of gluon coupling
constants $g_0$ and $g_1$ is similar to the S-wave baryon ones [38].

The estimation of theoretical error on the S-wave hexaquarks masses is
$1\, MeV$. This results was obtained by the choice of model parameters.
We predict the deuteron state as the mix of S- and D-wave contributions.
($NN_{SIJ=001}$ with the mass $M=1865\, MeV$ and $\Delta\Delta_{SIJ=001}$
with the mass $M=1834\, MeV$). The experimental value of deuteron mass is
$M=1876\, MeV$. In the cases of the $NN$, $\Delta\Delta$, $N\Delta$ systems
the Pauli principle requires that $(-1)^{L+I+J}=(-1)$, where $L$ the orbital
moment, $I$ isospin, $J$ spin of state are respectively. The wave function
of dibaryon must be antisymmetric for the permutation of all quarks. If we
consider the generalized Pauli rule for the wave function of dihyperons,
we must add the strangeness contribution to the isospin $I+\frac{S}{2}$.
Then we obtain the formula $(-1)^{L+I+\frac{S}{2}+J}=(-1)$. This rule allows
us to suggest the classification of dibaryons with the certain strangeness,
isospin and spin-parity (Tables I and II). We predict the degeneracy of the
some states. The contributions of subamplitudes to the hexaquark amplitude
are shown in the Appendix I (for example, $\Lambda\Lambda_{SIJ=-200}$).
The nonstrange dibaryon with the isospin $I=1$ and the spin-parity
$J^P=0^+$ is absent.

The states $N\Delta$ and $\Delta\Delta$ with the isospin $I=1$ and the
spin-parity $J^P=2^+$ possess the mass $M=2020\, MeV$. For the $N\Delta$
and $\Delta\Delta$ with the isospin $I=2$ and the spin-parity $J^P=1^+$
we obtained the mass $M=1984\, MeV$. For the state $\Delta\Delta$ with
the isospin $I=3$ and the spin-parity $J^P=0^+, \, 2^+$ ($M=2379\, MeV$)
the degeneracy is predicted. It is shown in the Table I.

The results for the strange sector of model are given in Table I and II.

\vskip2ex
{\bf IV. Conclusion.}
\vskip2ex

The dibaryon physics can be very delicate [39 -- 41]. The deuteron channel
is not a channel with strong attraction in any baryon interaction model.
If the deuteron had not been found experimentally, it seems highly unlikely
that any model would have been able to predict it to be a stable dibaryon.

The H-particle ($SIJ=-200$) is a six quark state consisting mainly of
octet-baryons, similar to the deuteron and one can find only a weak
attraction in the model [41]. Hence, a qualitative analysis is insufficient
to judge whether or not the H-particle is strong interaction stable.
Systematically, the authors find that a strong attraction develops only
in decuplet-decuplet channels and a mild attraction in octet-decuplet
channels [41]. Moreover, in the H-particle case, the channel coupling effect
may even be more important than the deuteron case. In fact, it is bound
without taking coupled channels into account. Besides the binding energy
of the H, an interesting question regarding the H is its compactness, i.e.
whether the H is a compact 6-quark object or a loosely bound
$\Lambda\Lambda$ state.

For systems with strangeness $S=-3$, Pang et al. have calculated the state
$N\Omega(SIJ=-3 \frac{1}{2} 2)$, which was shown to be midly attractive,
with energy below $\Lambda\Xi\pi$ threshold.

They have carried out a dynamical channel coupling calculation to examine
this state futher. The $N\Omega$, $\Lambda\Xi^*$, $\Xi\Sigma^*$,
$\Sigma\Xi^*$ $\Xi^*\Sigma^*$ channels are all included. The authors find
this to be a compact six quark state [41].

For systems with $S=-4$, with the quantum numbers $SIJ=-410$ as an
example, the lowest mass channel is composed of two
octet baryons from the same isodublet. The result shows that the system
with $S=-4$, $I=1$, $J=0$ is unbound, even when the $\Xi^*\Xi^*$ and
$\Sigma^*\Omega$ channel couplings are taken into account.

For comparison, Pang et al. have also calculated the $SIJ=-401$ state.
The $\Xi\Xi$, $\Xi\Xi^*$, $\Lambda\Omega$ and $\Xi^*\Xi^*$ coupling
channels are included. The result is very similar to the $SIJ=-410$,
i.e. they do not find a bound state in this channel.

For the systems with $S=-5$, Pang et al. take the $SIJ=-5 \frac{1}{2} 0$
state as an example. This state is interesting as a Pauli principle
favored state. If only two-baryon S-wave channels are taken into account,
there is only one channel for this state. The calculation shows [41] that
the contribution of the kinetic energy term, due to quark exchange and
delocalization effects, contributes strongly towards the formation of a
bound state. However, the one-gluon-exchange interaction largely compensates
for this attraction. Pang et al. conclude that this state is not a good
candidate for a dibaryon resonance search due to its small binding.

In the paper [41], Pang et al. would present a systematic study of possible
candidates of S-wave baryon-baryon bound states.

The H-particle, $N\Omega$-state and di-$\Omega$ may be strong interaction
stable. Up to now, these three interesting candidates of dibaryons are still
not found or confirmed by experiments. It seems that one should go beyond
these candidates and should search the possible candidates in a wider
region, especially the systems with multi-strangeness, in terms of a more
reliable model.

In our model the deuteron consist of the $\Delta\Delta$, $NN$ contributions
(the strangeness $S=0$, the isospin $I=0$, spin-parity $J^P=1^+$, the
quark content is $uuuddd$). The H-particle ($SIJ=-200$) content includes
$N\Xi$, $\Sigma^*\Sigma^*$, $\Sigma\Sigma$, $\Lambda\Lambda$.

For the systems with strangeness $S=-3$ ($N\Omega$ $SIJ=-3 \frac{1}{2} 2$)
the $N\Omega$, $\Lambda\Xi^*$, $\Xi\Sigma^*$, $\Sigma\Xi^*$, $\Xi^*\Sigma^*$
channels are included.

For the di-$\Omega$ state we consider the strangeness $S=-6$ ($SIJ=-600$).

The gluon coupling constants in our model is determined by the masses of
the H-particle and di-$\Omega$ state (Table I and II).

We considered 39 dibaryons, calculated the masses these states with the
strangeness $S=0, -1, -2, -3, -4, -5, -6$ and the spin-parity
$J^P=0^+, 1^+, 2^+$.

In our paper the dynamics of quark interactions in defined by the
Ghew-Mandelstam functions (Table III). We include only three parameters:
the cutoff $\Lambda$, gluon coupling constants $g_0$, $g_1$. The
relativistic six-body approach gives rise to the dynamical mixing of the
six-quark amplitudes and the dibaryon amplitudes. We calculated the masses
of two groups of dibaryons (Table I and II), which similar to the results
of other papers [12, 38 -- 40]. In our paper the relativistic description
of six particles amplitudes of S-wave dibaryons are considered. We use
only three parameters for the calculations of 39 dibaryon masses. The
interesting research is the consideration of the $qqqqqQ$ states with $Q$
a heavy quark ($Q=c, b$).

\vskip2.0ex
{\bf Acknowledgments.}
\vskip2.0ex

The authors would like to thank T. Barnes and C.-Y. Wong for useful
discussions. The work was carried with the support of the Russian Ministry
of Education (grant 2.1.1.68.26).

\newpage

{\bf Appendix I. The reduced amplitudes of dibaryons $NN_{SIJ=001}$,
$\Delta\Delta_{SIJ=001}$, $\Omega\Omega_{SIJ=-600}$,
$\Lambda\Lambda_{SIJ=-200}$, $N\Omega_{SIJ=-3 \frac{1}{2} 2}$ .}
\vskip2.0ex

$NN_{SIJ=001}$:

\begin{eqnarray}
\alpha_1^{1^{uu}}&=&\lambda+2\, \alpha_1^{1^{uu}} I_1(1^{uu}1^{uu})
+6\, \alpha_1^{0^{ud}} I_1(1^{uu}0^{ud})
+6\, \alpha_2^{1^{uu}0^{ud}} I_2(1^{uu}1^{uu}0^{ud})
+6\, \alpha_2^{0^{ud}0^{ud}} I_2(1^{uu}0^{ud}0^{ud})
\nonumber\\
&&\nonumber\\
\alpha_1^{1^{dd}}&=&\lambda+2\, \alpha_1^{1^{dd}} I_1(1^{dd}1^{dd})
+6\, \alpha_1^{0^{ud}} I_1(1^{dd}0^{ud})
+6\, \alpha_2^{1^{dd}0^{ud}} I_2(1^{dd}1^{dd}0^{ud})
+6\, \alpha_2^{0^{ud}0^{ud}} I_2(1^{dd}0^{ud}0^{ud})
\nonumber\\
&&\nonumber\\
\alpha_1^{0^{ud}}&=&\lambda+2\, \alpha_1^{1^{uu}} I_1(0^{ud}1^{uu})
+2\, \alpha_1^{1^{dd}} I_1(0^{ud}1^{dd})
+4\, \alpha_1^{0^{ud}} I_1(0^{ud}0^{ud})
+4\, \alpha_2^{1^{uu}1^{dd}} I_2(0^{ud}1^{uu}1^{dd})
\nonumber\\
&&\nonumber\\
&+&2\, \alpha_2^{1^{uu}0^{ud}} I_2(0^{ud}1^{uu}0^{ud})
+2\, \alpha_2^{1^{dd}0^{ud}} I_2(0^{ud}0^{ud}1^{dd})
+4\, \alpha_2^{0^{ud}0^{ud}} I_2(0^{ud}0^{ud}0^{ud})
\nonumber\\
&&\nonumber\\
\alpha_2^{1^{uu}1^{dd}}&=&\lambda
+\alpha_1^{0^{ud}} (4\, I_3(1^{uu}1^{dd}0^{ud})+2\, I_4(1^{uu}1^{dd}0^{ud})
+2\, I_4(1^{dd}1^{uu}0^{ud}))
+2\, \alpha_2^{1^{uu}0^{ud}} I_5(1^{uu}1^{dd}1^{uu}0^{ud})
\nonumber\\
&&\nonumber\\
&+&2\, \alpha_2^{1^{dd}0^{ud}} I_5(1^{dd}1^{uu}0^{ud}1^{dd})
+\alpha_2^{0^{ud}0^{ud}} (4\, I_6(1^{uu}1^{dd}0^{ud}0^{ud})
+4\, I_7(1^{uu}1^{dd}0^{ud}0^{ud})
\nonumber\\
&&\nonumber\\
&+&4\, I_7(1^{dd}1^{uu}0^{ud}0^{ud}))
\nonumber\\
&&\nonumber\\
\alpha_2^{1^{uu}0^{ud}}&=&\lambda
+2\, \alpha_1^{1^{uu}} I_3(1^{uu}0^{ud}1^{uu})
+2\, \alpha_1^{1^{dd}} I_4(0^{ud}1^{uu}1^{dd})
+\alpha_1^{0^{ud}} (2\, I_3(1^{uu}0^{ud}0^{ud})+4\, I_4(1^{uu}0^{ud}0^{ud})
\nonumber\\
&&\nonumber\\
&+&2\, I_4(0^{ud}1^{uu}0^{ud}))
+4\, \alpha_2^{1^{uu}1^{dd}} I_7(1^{uu}0^{ud}1^{uu}1^{dd})
+4\, \alpha_2^{1^{uu}0^{ud}} I_7(0^{ud}1^{uu}1^{uu}0^{ud})
\nonumber\\
&&\nonumber\\
&+&\alpha_2^{1^{dd}0^{ud}} (2\, I_5(0^{ud}1^{uu}0^{ud}1^{dd})
+4\, I_6(1^{uu}0^{ud}0^{ud}1^{dd}))
+\alpha_2^{0^{ud}0^{ud}} (2\, I_5(1^{uu}0^{ud}0^{ud}0^{ud})
\nonumber\\
&&\nonumber\\
&+&4\, I_6(1^{uu}0^{ud}0^{ud}0^{ud})
+4\, I_7(1^{uu}0^{ud}0^{ud}0^{ud})+4\, I_7(0^{ud}1^{uu}0^{ud}0^{ud}))
\nonumber\\
&&\nonumber\\
&+&4\, \alpha_3^{1^{uu}1^{dd}0^{ud}} I_8(1^{uu}0^{ud}0^{ud}1^{uu}1^{dd})
\nonumber\\
&&\nonumber\\
\alpha_2^{1^{dd}0^{ud}}&=&\lambda
+2\, \alpha_1^{1^{uu}} I_4(0^{ud}1^{dd}1^{uu})
+2\, \alpha_1^{1^{dd}} I_3(1^{dd}0^{ud}1^{dd})
+\alpha_1^{0^{ud}} (2\, I_3(1^{dd}0^{ud}0^{ud})+4\, I_4(1^{dd}0^{ud}0^{ud})
\nonumber\\
&&\nonumber\\
&+&2\, I_4(0^{ud}1^{dd}0^{ud}))
+4\, \alpha_2^{1^{uu}1^{dd}} I_7(1^{dd}0^{ud}1^{dd}1^{uu})
+\alpha_2^{1^{uu}0^{ud}} (2\, I_5(0^{ud}1^{dd}0^{ud}1^{uu})
\nonumber\\
&&\nonumber\\
&+&4\, I_6(1^{dd}0^{ud}0^{ud}1^{uu}))
+4\, \alpha_2^{1^{dd}0^{ud}} I_7(0^{ud}1^{dd}1^{dd}0^{ud})
+\alpha_2^{0^{ud}0^{ud}} (2\, I_5(1^{dd}0^{ud}0^{ud}0^{ud})
\nonumber\\
&&\nonumber\\
&+&4\, I_6(1^{dd}0^{ud}0^{ud}0^{ud})+4\, I_7(1^{dd}0^{ud}0^{ud}0^{ud})
+4\, I_7(0^{ud}1^{dd}0^{ud}0^{ud}))
\nonumber\\
&&\nonumber\\
&+&4\, \alpha_3^{1^{uu}1^{dd}0^{ud}} I_8(1^{dd}0^{ud}0^{ud}1^{dd}1^{uu})
\nonumber\\
&&\nonumber\\
\alpha_2^{0^{ud}0^{ud}}&=&\lambda+\alpha_1^{1^{uu}} (I_3(0^{ud}0^{ud}1^{uu})
+2\, I_4(0^{ud}0^{ud}1^{uu}))
+\alpha_1^{1^{dd}} (I_3(0^{ud}0^{ud}1^{dd})+2\, I_4(0^{ud}0^{ud}1^{dd}))
\nonumber\\
&&\nonumber\\
&+&\alpha_1^{0^{ud}} (2\, I_3(0^{ud}0^{ud}0^{ud})
+4\, I_4(0^{ud}0^{ud}0^{ud}))
+\alpha_2^{1^{uu}1^{dd}} (2\, I_5(0^{ud}0^{ud}1^{uu}1^{dd})
+2\, I_6(0^{ud}0^{ud}1^{uu}1^{dd})
\nonumber\\
&&\nonumber\\
&+&2\, I_7(0^{ud}0^{ud}1^{dd}1^{uu})+2\, I_7(0^{ud}0^{ud}1^{uu}1^{dd}))
+\alpha_2^{1^{uu}0^{ud}} (2\, I_6(0^{ud}0^{ud}1^{uu}0^{ud})
+2\, I_7(0^{ud}0^{ud}1^{uu}0^{ud})
\nonumber\\
&&\nonumber\\
&+&2\, I_7(0^{ud}0^{ud}0^{ud}1^{uu}))
+\alpha_2^{1^{dd}0^{ud}} (2\, I_6(0^{ud}0^{ud}0^{ud}1^{dd})
+2\, I_7(0^{ud}0^{ud}0^{ud}1^{dd})+2\, I_7(0^{ud}0^{ud}1^{dd}0^{ud}))
\nonumber\\
&&\nonumber\\
&+&\alpha_2^{0^{ud}0^{ud}} (2\, I_5(0^{ud}0^{ud}0^{ud}0^{ud})
+2\, I_6(0^{ud}0^{ud}0^{ud}0^{ud})+4\, I_7(0^{ud}0^{ud}0^{ud}0^{ud}))
\nonumber\\
&&\nonumber\\
&+&\alpha_3^{1^{uu}1^{dd}0^{ud}} (2\, I_8(0^{ud}0^{ud}1^{uu}0^{ud}1^{dd})
+2\, I_8(0^{ud}0^{ud}1^{uu}1^{dd}0^{ud})
+2\, I_8(0^{ud}0^{ud}0^{ud}1^{uu}1^{dd}))
\nonumber\\
&&\nonumber\\
\alpha_3^{1^{uu}1^{dd}0^{ud}}&=&\lambda
+2\, \alpha_1^{1^{uu}} I_9(1^{uu}0^{ud}1^{dd}1^{uu})
+2\, \alpha_1^{1^{dd}} I_9(1^{dd}0^{ud}1^{uu}1^{dd})
+\alpha_1^{0^{ud}} (4\, I_9(1^{uu}1^{dd}0^{ud}0^{ud})
\nonumber\\
&&\nonumber\\
&+&2\, I_9(1^{uu}0^{ud}1^{dd}0^{ud})
+2\, I_9(1^{dd}0^{ud}10^{ud}))
+4\, \alpha_2^{1^{uu}1^{dd}} I_{10}(1^{uu}0^{ud}1^{dd}1^{uu}1^{dd})
\nonumber\\
&&\nonumber\\
&+&4\, \alpha_2^{1^{uu}0^{ud}} I_{10}(1^{dd}1^{uu}0^{ud}0^{ud}1^{uu})
+4\, \alpha_2^{1^{dd}0^{ud}} I_{10}(1^{uu}1^{dd}0^{ud}0^{ud}1^{dd})
\nonumber\\
&&\nonumber\\
&+&\alpha_2^{0^{ud}0^{ud}} (4\, I_{10}(1^{uu}1^{dd}0^{ud}0^{ud}0^{ud})
+4\, I_{10}(1^{dd}1^{uu}0^{ud}0^{ud}0^{ud})
+4\, I_{10}(1^{uu}0^{ud}1^{dd}0^{ud}0^{ud}))\nonumber
\end{eqnarray}

\vskip5ex

$\Delta\Delta_{SIJ=001}$:

\begin{eqnarray}
\alpha_1^{1^{uu}}&=&\lambda+2\, \alpha_1^{1^{uu}} I_1(1^{uu}1^{uu})
+6\, \alpha_1^{0^{ud}} I_1(1^{uu}0^{ud})
+6\, \alpha_2^{1^{uu}0^{ud}} I_2(1^{uu}1^{uu}0^{ud})
+6\, \alpha_2^{0^{ud}0^{ud}} I_2(1^{uu}0^{ud}0^{ud})
\nonumber\\
&&\nonumber\\
\alpha_1^{1^{dd}}&=&\lambda+2\, \alpha_1^{1^{dd}} I_1(1^{dd}1^{dd})
+6\, \alpha_1^{0^{ud}} I_1(1^{dd}0^{ud})
+6\, \alpha_2^{1^{dd}0^{ud}} I_2(1^{dd}1^{dd}0^{ud})
+6\, \alpha_2^{0^{ud}0^{ud}} I_2(1^{dd}0^{ud}0^{ud})
\nonumber\\
&&\nonumber\\
\alpha_1^{0^{ud}}&=&\lambda+2\, \alpha_1^{1^{uu}} I_1(0^{ud}1^{uu})
+2\, \alpha_1^{1^{dd}} I_1(0^{ud}1^{dd})
+4\, \alpha_1^{0^{ud}} I_1(0^{ud}0^{ud})
+4\, \alpha_2^{1^{uu}1^{dd}} I_2(0^{ud}1^{uu}1^{dd})
\nonumber\\
&&\nonumber\\
&+&2\, \alpha_2^{1^{uu}0^{ud}} I_2(0^{ud}1^{uu}0^{ud})
+2\, \alpha_2^{1^{dd}0^{ud}} I_2(0^{ud}0^{ud}1^{dd})
+4\, \alpha_2^{0^{ud}0^{ud}} I_2(0^{ud}0^{ud}0^{ud})
\nonumber\\
&&\nonumber\\
\alpha_2^{1^{uu}1^{dd}}&=&\lambda
+2\, \alpha_1^{1^{uu}} I_4(1^{uu}1^{dd}1^{uu})
+2\, \alpha_1^{1^{dd}} I_4(1^{dd}1^{uu}1^{dd})
+\alpha_1^{0^{ud}} (4\, I_3(1^{uu}1^{dd}0^{ud})+2\, I_4(1^{uu}1^{dd}0^{ud})
\nonumber\\
&&\nonumber\\
&+&2\, I_4(1^{dd}1^{uu}0^{ud}))
+4\, \alpha_2^{1^{uu}1^{dd}} I_6(1^{uu}1^{dd}1^{uu}1^{dd})
+\alpha_2^{1^{uu}0^{ud}} (2\, I_5(1^{uu}1^{dd}1^{uu}0^{ud})
\nonumber\\
&&\nonumber\\
&+&4\, I_7(1^{dd}1^{uu}0^{ud}1^{uu}))
+\alpha_2^{1^{dd}0^{ud}} (2\, I_5(1^{dd}1^{uu}0^{ud}1^{dd})
+4\, I_7(1^{uu}1^{dd}0^{ud}1^{dd}))
\nonumber\\
&&\nonumber\\
&+&\alpha_2^{0^{ud}0^{ud}} (4\, I_6(1^{uu}1^{dd}0^{ud}0^{ud})
+4\, I_7(1^{uu}1^{dd}0^{ud}0^{ud})+4\, I_7(1^{dd}1^{uu}0^{ud}0^{ud}))
\nonumber\\
&&\nonumber\\
&+&4\, \alpha_3^{1^{uu}1^{dd}0^{ud}} I_8(1^{uu}1^{dd}1^{uu}0^{ud}1^{dd})
\nonumber\\
&&\nonumber\\
\alpha_2^{1^{uu}0^{ud}}&=&\lambda
+2\, \alpha_1^{1^{uu}} I_3(1^{uu}0^{ud}1^{uu})
+2\, \alpha_1^{1^{dd}} I_4(0^{ud}1^{uu}1^{dd})
+\alpha_1^{0^{ud}} (2\, I_3(1^{uu}0^{ud}0^{ud})+4\, I_4(1^{uu}0^{ud}0^{ud})
\nonumber\\
&&\nonumber\\
&+&2\, I_4(0^{ud}1^{uu}0^{ud}))
+4\, \alpha_2^{1^{uu}1^{dd}} I_7(1^{uu}0^{ud}1^{uu}1^{dd})
+4\, \alpha_2^{1^{uu}0^{ud}} I_7(0^{ud}1^{uu}1^{uu}0^{ud})
\nonumber\\
&&\nonumber\\
&+&\alpha_2^{1^{dd}0^{ud}} (2\, I_5(0^{ud}1^{uu}0^{ud}1^{dd})
+4\, I_6(1^{uu}0^{ud}0^{ud}1^{dd}))
+\alpha_2^{0^{ud}0^{ud}} (2\, I_5(1^{uu}0^{ud}0^{ud}0^{ud})
\nonumber\\
&&\nonumber\\
&+&4\, I_6(1^{uu}0^{ud}0^{ud}0^{ud})
+4\, I_7(1^{uu}0^{ud}0^{ud}0^{ud})+4\, I_7(0^{ud}1^{uu}0^{ud}0^{ud}))
\nonumber\\
&&\nonumber\\
&+&4\, \alpha_3^{1^{uu}1^{dd}0^{ud}} I_8(1^{uu}0^{ud}0^{ud}1^{uu}1^{dd})
\nonumber\\
&&\nonumber\\
\alpha_2^{1^{dd}0^{ud}}&=&\lambda
+2\, \alpha_1^{1^{uu}} I_4(0^{ud}1^{dd}1^{uu})
+2\, \alpha_1^{1^{dd}} I_3(1^{dd}0^{ud}1^{dd})
+\alpha_1^{0^{ud}} (2\, I_3(1^{dd}0^{ud}0^{ud})+4\, I_4(1^{dd}0^{ud}0^{ud})
\nonumber\\
&&\nonumber\\
&+&2\, I_4(0^{ud}1^{dd}0^{ud}))
+4\, \alpha_2^{1^{uu}1^{dd}} I_7(1^{dd}0^{ud}1^{dd}1^{uu})
+\alpha_2^{1^{uu}0^{ud}} (2\, I_5(0^{ud}1^{dd}0^{ud}1^{uu})
\nonumber\\
&&\nonumber\\
&+&4\, I_6(1^{dd}0^{ud}0^{ud}1^{uu}))
+4\, \alpha_2^{1^{dd}0^{ud}} I_7(0^{ud}1^{dd}1^{dd}0^{ud})
+\alpha_2^{0^{ud}0^{ud}} (2\, I_5(1^{dd}0^{ud}0^{ud}0^{ud})
\nonumber\\
&&\nonumber\\
&+&4\, I_6(1^{dd}0^{ud}0^{ud}0^{ud})+4\, I_7(1^{dd}0^{ud}0^{ud}0^{ud})
+4\, I_7(0^{ud}1^{dd}0^{ud}0^{ud}))
\nonumber\\
&&\nonumber\\
&+&4\, \alpha_3^{1^{uu}1^{dd}0^{ud}} I_8(1^{dd}0^{ud}0^{ud}1^{dd}1^{uu})
\nonumber\\
&&\nonumber\\
\alpha_2^{0^{ud}0^{ud}}&=&\lambda+\alpha_1^{1^{uu}} (I_3(0^{ud}0^{ud}1^{uu})
+2\, I_4(0^{ud}0^{ud}1^{uu}))
+\alpha_1^{1^{dd}} (I_3(0^{ud}0^{ud}1^{dd})+2\, I_4(0^{ud}0^{ud}1^{dd}))
\nonumber\\
&&\nonumber\\
&+&\alpha_1^{0^{ud}} (2\, I_3(0^{ud}0^{ud}0^{ud})
+4\, I_4(0^{ud}0^{ud}0^{ud}))
+\alpha_2^{1^{uu}1^{dd}} (2\, I_5(0^{ud}0^{ud}1^{uu}1^{dd})
+2\, I_6(0^{ud}0^{ud}1^{uu}1^{dd})
\nonumber\\
&&\nonumber\\
&+&2\, I_7(0^{ud}0^{ud}1^{dd}1^{uu})+2\, I_7(0^{ud}0^{ud}1^{uu}1^{dd}))
+\alpha_2^{1^{uu}0^{ud}} (2\, I_6(0^{ud}0^{ud}1^{uu}0^{ud})
+2\, I_7(0^{ud}0^{ud}1^{uu}0^{ud})
\nonumber\\
&&\nonumber\\
&+&2\, I_7(0^{ud}0^{ud}0^{ud}1^{uu}))
+\alpha_2^{1^{dd}0^{ud}} (2\, I_6(0^{ud}0^{ud}0^{ud}1^{dd})
+2\, I_7(0^{ud}0^{ud}0^{ud}1^{dd})+2\, I_7(0^{ud}0^{ud}1^{dd}0^{ud}))
\nonumber\\
&&\nonumber\\
&+&\alpha_2^{0^{ud}0^{ud}} (2\, I_5(0^{ud}0^{ud}0^{ud}0^{ud})
+2\, I_6(0^{ud}0^{ud}0^{ud}0^{ud})+4\, I_7(0^{ud}0^{ud}0^{ud}0^{ud}))
\nonumber\\
&&\nonumber\\
&+&\alpha_3^{1^{uu}1^{dd}0^{ud}} (2\, I_8(0^{ud}0^{ud}1^{uu}0^{ud}1^{dd})
+2\, I_8(0^{ud}0^{ud}1^{uu}1^{dd}0^{ud})
+2\, I_8(0^{ud}0^{ud}0^{ud}1^{uu}1^{dd}))
\nonumber\\
&&\nonumber\\
\alpha_3^{1^{uu}1^{dd}0^{ud}}&=&\lambda
+2\, \alpha_1^{1^{uu}} I_9(1^{uu}0^{ud}1^{dd}1^{uu})
+2\, \alpha_1^{1^{dd}} I_9(1^{dd}0^{ud}1^{uu}1^{dd})
+\alpha_1^{0^{ud}} (4\, I_9(1^{uu}1^{dd}0^{ud}0^{ud})
\nonumber\\
&&\nonumber\\
&+&2\, I_9(1^{uu}0^{ud}1^{dd}0^{ud})
+2\, I_9(1^{dd}0^{ud}10^{ud}))
+4\, \alpha_2^{1^{uu}1^{dd}} I_{10}(1^{uu}0^{ud}1^{dd}1^{uu}1^{dd})
\nonumber\\
&&\nonumber\\
&+&4\, \alpha_2^{1^{uu}0^{ud}} I_{10}(1^{dd}1^{uu}0^{ud}0^{ud}1^{uu})
+4\, \alpha_2^{1^{dd}0^{ud}} I_{10}(1^{uu}1^{dd}0^{ud}0^{ud}1^{dd})
\nonumber\\
&&\nonumber\\
&+&\alpha_2^{0^{ud}0^{ud}} (4\, I_{10}(1^{uu}1^{dd}0^{ud}0^{ud}0^{ud})
+4\, I_{10}(1^{dd}1^{uu}0^{ud}0^{ud}0^{ud})
+4\, I_{10}(1^{uu}0^{ud}1^{dd}0^{ud}0^{ud}))
\nonumber
\end{eqnarray}

\newpage

$\Omega\Omega_{SIJ=-600}$:

\begin{eqnarray}
\alpha_1^{1^{ss}} &=&\lambda+8\, \alpha_1^{1^{ss}} I_1(1^{ss}1^{ss})
+12\, \alpha_2^{1^{ss}1^{ss}} I_2(1^{ss}1^{ss}1^{ss})
\nonumber\\
&&\nonumber\\
\alpha_2^{1^{ss}1^{ss}} &=&\lambda
+\alpha_1^{1^{ss}} (4\, I_3(1^{ss}1^{ss}1^{ss})+8\, I_4(1^{ss}1^{ss}1^{ss}))
+\alpha_2^{1^{ss}1^{ss}} (4\, I_5(1^{ss}1^{ss}1^{ss}1^{ss})
\nonumber\\
&&\nonumber\\
&+&8\, I_6(1^{ss}1^{ss}1^{ss}1^{ss})+16\, I_7(1^{ss}1^{ss}1^{ss}1^{ss}))
+8\, \alpha_3^{1^{ss}1^{ss}1^{ss}} I_8(1^{ss}1^{ss}1^{ss}1^{ss}1^{ss})
\nonumber\\
&&\nonumber\\
\alpha_3^{1^{ss}1^{ss}1^{ss}} &=&\lambda
+12\, \alpha_1^{1^{ss}} I_9(1^{ss}1^{ss}1^{ss}1^{ss})
+24\, \alpha_2^{1^{ss}1^{ss}} I_{10}(1^{ss}1^{ss}1^{ss}1^{ss}1^{ss})
\nonumber
\end{eqnarray}

\vskip5ex

$\Lambda\Lambda_{SIJ=-200}$:

\begin{eqnarray}
\alpha_1^{1^{uu}}&=&\lambda+4\, \alpha_1^{0^{ud}} I_1(1^{uu}0^{ud})
+4\, \alpha_1^{0^{us}} I_1(1^{uu}0^{us})
+2\, \alpha_2^{0^{ud}0^{ud}} I_2(1^{uu}0^{ud}0^{ud})
+8\, \alpha_2^{0^{ud}0^{us}} I_2(1^{uu}0^{ud}0^{us})
\nonumber\\
&&\nonumber\\
&+&2\, \alpha_2^{0^{us}0^{us}} I_2(1^{uu}0^{us}0^{us})
\nonumber\\
&&\nonumber\\
\alpha_1^{1^{dd}}&=&\lambda+4\, \alpha_1^{0^{ud}} I_1(1^{dd}0^{ud})
+4\, \alpha_1^{0^{ds}} I_1(1^{dd}0^{ds})
+2\, \alpha_2^{0^{ud}0^{ud}} I_2(1^{dd}0^{ud}0^{ud})
+8\, \alpha_2^{0^{ud}0^{ds}} I_2(1^{dd}0^{ud}0^{ds})
\nonumber\\
&&\nonumber\\
&+&2\, \alpha_2^{0^{ds}0^{ds}} I_2(1^{dd}0^{ds}0^{ds})
\nonumber\\
&&\nonumber\\
\alpha_1^{1^{ss}}&=&\lambda+4\, \alpha_1^{0^{us}} I_1(1^{ss}0^{us})
+4\, \alpha_1^{0^{ds}} I_1(1^{ss}0^{ds})
+2\, \alpha_2^{0^{us}0^{us}} I_2(1^{ss}0^{us}0^{us})
+8\, \alpha_2^{0^{us}0^{ds}} I_2(1^{ss}0^{us}0^{ds})
\nonumber\\
&&\nonumber\\
&+&2\, \alpha_2^{0^{ds}0^{ds}} I_2(1^{ss}0^{ds}0^{ds})
\nonumber\\
&&\nonumber\\
\alpha_1^{0^{ud}}&=&\lambda+\alpha_1^{1^{uu}} I_1(0^{ud}1^{uu})
+\alpha_1^{1^{dd}} I_1(0^{ud}1^{dd})+2\, \alpha_1^{0^{ud}} I_1(0^{ud}0^{ud})
+2\, \alpha_1^{0^{us}} I_1(0^{ud}0^{us})
\nonumber\\
&&\nonumber\\
&+&2\, \alpha_1^{0^{ds}} I_1(0^{ud}0^{ds})
+\alpha_2^{0^{ud}0^{ud}} I_2(0^{ud}0^{ud}0^{ud})
+2\, \alpha_2^{0^{ud}0^{us}} I_2(0^{ud}0^{ud}0^{us})
+2\, \alpha_2^{0^{ud}0^{ds}} I_2(0^{ud}0^{ud}0^{ds})
\nonumber\\
&&\nonumber\\
&+&2\, \alpha_2^{0^{us}0^{ds}} I_2(0^{ud}0^{us}0^{ds})
\nonumber\\
&&\nonumber\\
\alpha_1^{0^{us}}&=&\lambda+\alpha_1^{1^{uu}} I_1(0^{us}1^{uu})
+\alpha_1^{1^{ss}} I_1(0^{us}1^{ss})+2\, \alpha_1^{0^{ud}} I_1(0^{us}0^{ud})
+2\, \alpha_1^{0^{us}} I_1(0^{us}0^{us})
\nonumber\\
&&\nonumber\\
&+&2\, \alpha_1^{0^{ds}} I_1(0^{us}0^{ds})
+2\, \alpha_2^{0^{ud}0^{us}} I_2(0^{us}0^{ud}0^{us})
+2\, \alpha_2^{0^{ud}0^{ds}} I_2(0^{us}0^{ud}0^{ds})
+\alpha_2^{0^{us}0^{us}} I_2(0^{us}0^{us}0^{us})
\nonumber\\
&&\nonumber\\
&+&2\, \alpha_2^{0^{us}0^{ds}} I_2(0^{us}0^{us}0^{ds})
\nonumber\\
&&\nonumber\\
\alpha_1^{0^{ds}}&=&\lambda+\alpha_1^{1^{dd}} I_1(0^{ds}1^{dd})
+\alpha_1^{1^{ss}} I_1(0^{ds}1^{ss})+2\, \alpha_1^{0^{ud}} I_1(0^{ds}0^{ud})
+2\, \alpha_1^{0^{us}} I_1(0^{ds}0^{us})
\nonumber\\
&&\nonumber\\
&+&2\, \alpha_1^{0^{ds}} I_1(0^{ds}0^{ds})
+2\, \alpha_2^{0^{ud}0^{us}} I_2(0^{ds}0^{ud}0^{us})
+2\, \alpha_2^{0^{ud}0^{ds}} I_2(0^{ds}0^{ud}0^{ds})
+2\, \alpha_2^{0^{us}0^{ds}} I_2(0^{ds}0^{us}0^{ds})
\nonumber\\
&&\nonumber\\
&+&\alpha_2^{0^{ds}0^{ds}} I_2(0^{ds}0^{ds}0^{ds})
\nonumber\\
&&\nonumber\\
\alpha_2^{0^{ud}0^{ud}}&=&\lambda+\alpha_1^{1^{uu}} I_3(0^{ud}0^{ud}1^{uu})
+\alpha_1^{1^{dd}} I_3(0^{ud}0^{ud}1^{dd})
+2\, \alpha_1^{0^{ud}} I_3(0^{ud}0^{ud}0^{ud})
+4\, \alpha_1^{0^{us}} I_4(0^{ud}0^{ud}0^{us})
\nonumber\\
&&\nonumber\\
&+&4\, \alpha_1^{0^{ds}} I_4(0^{ud}0^{ud}0^{ds})
+4\, \alpha_2^{0^{ud}0^{us}} I_7(0^{ud}0^{ud}0^{ud}0^{us})
+4\, \alpha_2^{0^{ud}0^{ds}} I_7(0^{ud}0^{ud}0^{ud}0^{ds})
\nonumber\\
&&\nonumber\\
&+&2\, \alpha_2^{0^{us}0^{us}} I_6(0^{ud}0^{ud}0^{us}0^{us})
+4\, \alpha_2^{0^{us}0^{ds}} (I_5(0^{ud}0^{ud}0^{us}0^{ds})
+I_6(0^{ud}0^{ud}0^{us}0^{ds}))
\nonumber\\
&&\nonumber\\
&+&2\, \alpha_2^{0^{ds}0^{ds}} I_6(0^{ud}0^{ud}0^{ds}0^{ds})
+4\, \alpha_3^{0^{ud}0^{us}0^{ds}} I_8(0^{ud}0^{ud}0^{us}0^{ud}0^{ds})
\nonumber\\
&&\nonumber\\
\alpha_2^{0^{ud}0^{us}}&=&\lambda+\alpha_1^{1^{uu}} I_3(0^{ud}0^{us}1^{uu})
+\alpha_1^{0^{ud}} (I_3(0^{ud}0^{us}0^{ud})+I_4(0^{us}0^{ud}0^{ud}))
+\alpha_1^{0^{us}} (I_3(0^{ud}0^{us}0^{us})
\nonumber\\
&&\nonumber\\
&+&I_4(0^{ud}0^{us}0^{us}))
+\alpha_1^{0^{ds}} (I_3(0^{ud}0^{us}0^{ds})+I_4(0^{ud}0^{us}0^{ds})
+I_4(0^{us}0^{ud}0^{ds}))
\nonumber\\
&&\nonumber\\
&+&\alpha_2^{0^{ud}0^{us}} (I_6(0^{ud}0^{us}0^{us}0^{ud})
+I_7(0^{ud}0^{us}0^{us}0^{ud})+I_7(0^{us}0^{ud}0^{ud}0^{us}))
+\alpha_2^{0^{ud}0^{ds}} (I_5(0^{ud}0^{us}0^{ud}0^{ds})
\nonumber\\
&&\nonumber\\
&+&I_6(0^{ud}0^{us}0^{ds}0^{ud})
+I_7(0^{ud}0^{us}0^{ud}0^{ds})+I_7(0^{ud}0^{us}0^{ds}0^{ud}))
+\alpha_2^{0^{us}0^{ds}} (I_5(0^{us}0^{ud}0^{us}0^{ds})
\nonumber\\
&&\nonumber\\
&+&I_6(0^{ud}0^{us}0^{us}0^{ds})
+I_7(0^{us}0^{ud}0^{us}0^{ds})+I_7(0^{us}0^{ud}0^{ds}0^{us}))
+\alpha_2^{0^{ds}0^{ds}} I_6(0^{ud}0^{us}0^{ds}0^{ds})
\nonumber\\
&&\nonumber\\
&+&\alpha_3^{0^{ud}0^{us}0^{ds}} (I_8(0^{ud}0^{us}0^{us}0^{ud}0^{ds})
+I_8(0^{ud}0^{us}0^{us}0^{ds}0^{ud})+I_8(0^{ud}0^{us}0^{ds}0^{us}0^{ud}))
\nonumber\\
&&\nonumber\\
\alpha_2^{0^{ud}0^{ds}}&=&\lambda+\alpha_1^{1^{dd}} I_3(0^{ud}0^{ds}1^{dd})
+\alpha_1^{0^{ud}} (I_3(0^{ud}0^{ds}0^{ud})+I_4(0^{ds}0^{ud}0^{ud}))
+\alpha_1^{0^{us}} (I_3(0^{ud}0^{ds}0^{us})
\nonumber\\
&&\nonumber\\
&+&I_4(0^{ud}0^{ds}0^{us})+I_4(0^{ds}0^{ud}0^{us}))
+\alpha_1^{0^{ds}} (I_3(0^{ud}0^{ds}0^{ds})+I_4(0^{ud}0^{ds}0^{ds}))
\nonumber\\
&&\nonumber\\
&+&\alpha_2^{0^{ud}0^{us}} (I_5(0^{ud}0^{ds}0^{ud}0^{us})
+I_6(0^{ud}0^{ds}0^{us}0^{ud})+I_7(0^{ud}0^{ds}0^{ud}0^{us})
+I_7(0^{ud}0^{ds}0^{us}0^{ud}))
\nonumber\\
&&\nonumber\\
&+&\alpha_2^{0^{ud}0^{ds}} (I_6(0^{ud}0^{ds}0^{ds}0^{ud})
+I_7(0^{ud}0^{ds}0^{ds}0^{ud})+I_7(0^{ds}0^{ud}0^{ud}0^{ds}))
+\alpha_2^{0^{us}0^{us}} I_6(0^{ud}0^{ds}0^{us}0^{us})
\nonumber\\
&&\nonumber\\
&+&\alpha_2^{0^{us}0^{ds}} (I_5(0^{ds}0^{ud}0^{us}0^{ds})
+I_6(0^{ud}0^{ds}0^{us}0^{ds})
+I_7(0^{ds}0^{ud}0^{us}0^{ds})+I_7(0^{ds}0^{ud}0^{ds}0^{us}))
\nonumber\\
&&\nonumber\\
&+&\alpha_3^{0^{ud}0^{us}0^{ds}} (I_8(0^{ud}0^{ds}0^{ds}0^{ud}0^{us})
+I_8(0^{ud}0^{ds}0^{us}0^{ds}0^{ud})+I_8(0^{ud}0^{ds}0^{ds}0^{us}0^{ud}))
\nonumber\\
&&\nonumber\\
\alpha_2^{0^{us}0^{us}}&=&\lambda+\alpha_1^{1^{uu}} I_3(0^{us}0^{us}1^{uu})
+\alpha_1^{1^{ss}} I_3(0^{us}0^{us}1^{ss})
+4\, \alpha_1^{0^{ud}} I_4(0^{us}0^{us}0^{ud})
+2\, \alpha_1^{0^{us}} I_3(0^{us}0^{us}0^{us})
\nonumber\\
&&\nonumber\\
&+&4\, \alpha_1^{0^{ds}} I_4(0^{us}0^{us}0^{ds})
+2\, \alpha_2^{0^{ud}0^{ud}} I_7(0^{us}0^{us}0^{ud}0^{ud})
+4\, \alpha_2^{0^{ud}0^{us}} I_7(0^{us}0^{us}0^{us}0^{ud})
\nonumber\\
&&\nonumber\\
&+&4\, \alpha_2^{0^{ud}0^{ds}} (I_5(0^{us}0^{us}0^{ud}0^{ds})
+I_6(0^{us}0^{us}0^{ud}0^{ds}))
+4\, \alpha_2^{0^{us}0^{ds}} I_7(0^{us}0^{us}0^{us}0^{ds})
\nonumber\\
&&\nonumber\\
&+&2\, \alpha_2^{0^{ds}0^{ds}} I_6(0^{us}0^{us}0^{ds}0^{ds})
+2\, \alpha_3^{0^{ud}0^{us}0^{ds}} I_8(0^{us}0^{us}0^{ud}0^{us}0^{ds})
\nonumber\\
&&\nonumber\\
\alpha_2^{0^{us}0^{ds}}&=&\lambda+\alpha_1^{1^{ss}} I_3(0^{us}0^{ds}1^{ss})
+\alpha_1^{0^{ud}} (I_3(0^{us}0^{ds}0^{ud})
+I_4(0^{us}0^{ds}0^{ud})+I_4(0^{ds}0^{us}0^{ud}))
\nonumber\\
&&\nonumber\\
&+&\alpha_1^{0^{us}} (I_3(0^{us}0^{ds}0^{us})+I_4(0^{ds}0^{us}0^{us}))
+\alpha_1^{0^{ds}} (I_3(0^{us}0^{ds}0^{ds})+I_4(0^{us}0^{ds}0^{ds}))
\nonumber\\
&&\nonumber\\
&+&\alpha_2^{0^{ud}0^{ud}} I_6(0^{us}0^{ds}0^{ud}0^{ud})
+\alpha_2^{0^{ud}0^{us}} (I_5(0^{us}0^{ds}0^{ud}0^{us})
+I_6(0^{us}0^{ds}0^{ud}0^{us})+I_7(0^{us}0^{ds}0^{ud}0^{us})
\nonumber\\
&&\nonumber\\
&+&I_7(0^{us}0^{ds}0^{us}0^{ud}))
+\alpha_2^{0^{ud}0^{ds}} (I_5(0^{ds}0^{us}0^{ud}0^{ds})
+I_6(0^{ds}0^{us}0^{ud}0^{ds})+I_7(0^{ds}0^{us}0^{ud}0^{ds})
\nonumber\\
&&\nonumber\\
&+&I_7(0^{ds}0^{us}0^{ds}0^{ud}))
+\alpha_2^{0^{us}0^{ds}} (I_6(0^{us}0^{ds}0^{ds}0^{us})
+I_7(0^{us}0^{ds}0^{ds}0^{us})+I_7(0^{ds}0^{us}0^{us}0^{ds}))
\nonumber\\
&&\nonumber\\
&+&\alpha_3^{0^{ud}0^{us}0^{ds}} (I_8(0^{us}0^{ds}0^{ds}0^{ud}0^{us})
+I_8(0^{us}0^{ds}0^{ud}0^{ds}0^{us})+I_8(0^{us}0^{ds}0^{ds}0^{us}0^{ud}))
\nonumber\\
&&\nonumber\\
\alpha_2^{0^{ds}0^{ds}}&=&\lambda+\alpha_1^{1^{dd}} I_3(0^{ds}0^{ds}1^{dd})
+\alpha_1^{1^{ss}} I_3(0^{ds}0^{ds}1^{ss})
+4\, \alpha_1^{0^{ud}} I_4(0^{ds}0^{ds}0^{ud})
+4\, \alpha_1^{0^{us}} I_4(0^{ds}0^{ds}0^{us})
\nonumber\\
&&\nonumber\\
&+&2\, \alpha_1^{0^{ds}} I_3(0^{ds}0^{ds}0^{ds})
+2\, \alpha_2^{0^{ud}0^{ud}} I_7(0^{ds}0^{ds}0^{ud}0^{ud})
+4\, \alpha_2^{0^{ud}0^{us}} (I_5(0^{ds}0^{ds}0^{ud}0^{us})
\nonumber\\
&&\nonumber\\
&+&I_6(0^{ds}0^{ds}0^{ud}0^{us}))
+4\, \alpha_2^{0^{ud}0^{ds}} I_7(0^{ds}0^{ds}0^{ds}0^{ud})
+2\, \alpha_2^{0^{us}0^{us}} I_6(0^{ds}0^{ds}0^{us}0^{us})
\nonumber\\
&&\nonumber\\
&+&4\, \alpha_2^{0^{us}0^{ds}} I_7(0^{ds}0^{ds}0^{ds}0^{us})
+2\, \alpha_3^{0^{ud}0^{us}0^{ds}} I_8(0^{ds}0^{ds}0^{ud}0^{ds}0^{us})
\nonumber\\
&&\nonumber\\
\alpha_3^{0^{ud}0^{us}0^{ds}}&=&\lambda
+\alpha_1^{1^{uu}} I_9(0^{ud}0^{us}0^{ds}1^{uu})
+\alpha_1^{1^{dd}} I_9(0^{ud}0^{ds}0^{us}1^{dd})
+\alpha_1^{1^{ss}} I_9(0^{us}0^{ds}0^{ud}1^{ss})
\nonumber\\
&&\nonumber\\
&+&\alpha_1^{0^{ud}} (I_9(0^{ud}0^{us}0^{ds}0^{ud})
+I_9(0^{ud}0^{ds}0^{us}0^{ud})+I_9(0^{us}0^{ds}0^{ud}0^{ud}))
+\alpha_1^{0^{us}} (I_9(0^{ud}0^{us}0^{ds}0^{us})
\nonumber\\
&&\nonumber\\
&+&I_9(0^{ud}0^{ds}0^{us}0^{us})+I_9(0^{us}0^{ds}0^{ud}0^{us}))
+\alpha_1^{0^{ds}} (I_9(0^{ud}0^{us}0^{ds}0^{ds})
+I_9(0^{ud}0^{ds}0^{us}0^{ds})
\nonumber\\
&&\nonumber\\
&+&I_9(0^{us}0^{ds}0^{ud}0^{ds}))
+\alpha_2^{0^{ud}0^{ud}} I_{10}(0^{us}0^{ud}0^{ds}0^{ud}0^{ud})
+\alpha_2^{0^{ud}0^{us}} (I_{10}(0^{ud}0^{ds}0^{us}0^{ud}0^{us})
\nonumber\\
&&\nonumber\\
&+&I_{10}(0^{us}0^{ud}0^{ds}0^{ud}0^{us})
+I_{10}(0^{us}0^{ds}0^{ud}0^{ud}0^{us})
+I_{10}(0^{ds}0^{us}0^{ud}0^{ud}0^{us}))
\nonumber\\
&&\nonumber\\
&+&\alpha_2^{0^{ud}0^{ds}} (I_{10}(0^{ud}0^{us}0^{ds}0^{ud}0^{ds})
+I_{10}(0^{ds}0^{ud}0^{us}0^{ud}0^{ds})
+I_{10}(0^{ds}0^{us}0^{ud}0^{ud}0^{ds})
\nonumber\\
&&\nonumber\\
&+&I_{10}(0^{us}0^{ds}0^{ud}0^{ud}0^{ds}))
+\alpha_2^{0^{us}0^{us}} I_{10}(0^{ud}0^{us}0^{ds}0^{us}0^{us})
+\alpha_2^{0^{us}0^{ds}} (I_{10}(0^{ud}0^{ds}0^{us}0^{us}0^{ds})
\nonumber\\
&&\nonumber\\
&+&I_{10}(0^{ds}0^{us}0^{ud}0^{us}0^{ds})
+I_{10}(0^{us}0^{ud}0^{ds}0^{us}0^{ds})
+I_{10}(0^{ds}0^{ud}0^{us}0^{us}0^{ds}))
\nonumber\\
&&\nonumber\\
&+&\alpha_2^{0^{ds}0^{ds}} I_{10}(0^{ud}0^{ds}0^{us}0^{ds}0^{ds})\nonumber
\end{eqnarray}

\newpage

$N\Omega_{SIJ=-3 \frac{1}{2} 2}$:

\begin{eqnarray}
\alpha_1^{1^{uu}}&=&\lambda+2\,\alpha_1^{0^{ud}} I_1(1^{uu}0^{ud})
+6\,\alpha_1^{0^{us}} I_1(1^{uu}0^{us})
\nonumber\\
&&\nonumber\\
\alpha_1^{1^{ss}}&=&\lambda+2\,\alpha_1^{1^{ss}} I_1(1^{ss}1^{ss})
+4\,\alpha_1^{0^{us}} I_1(1^{ss}0^{us})
+2\,\alpha_1^{0^{ds}} I_1(1^{ss}0^{ds})
\nonumber\\
&&\nonumber\\
\alpha_1^{0^{ud}}&=&\lambda+\alpha_1^{1^{uu}} I_1(0^{ud}1^{uu})
+\alpha_1^{0^{ud}} I_1(0^{ud}0^{ud})
+3\,\alpha_1^{0^{us}} I_1(0^{ud}0^{us})
+3\,\alpha_1^{0^{ds}} I_1(0^{ud}0^{ds})
\nonumber\\
&&\nonumber\\
\alpha_1^{0^{us}}&=&\lambda+\alpha_1^{1^{uu}} I_1(0^{us}1^{uu})
+2\,\alpha_1^{1^{ss}} I_1(0^{us}1^{ss})
+\alpha_1^{0^{ud}} I_1(0^{us}0^{ud})
+3\,\alpha_1^{0^{us}} I_1(0^{us}0^{us})
+\alpha_1^{0^{ds}} I_1(0^{us}0^{ds})
\nonumber\\
&&\nonumber\\
&+&2\,\alpha_2^{1^{uu}1^{ss}} I_2(0^{us}1^{uu}1^{ss})
+2\,\alpha_2^{1^{ss}0^{ud}} I_2(0^{us}0^{ud}1^{ss})
\nonumber\\
&&\nonumber\\
\alpha_1^{0^{ds}}&=&\lambda+2\,\alpha_1^{1^{ss}} I_1(0^{ds}1^{ss})
+2\,\alpha_1^{0^{ud}} I_1(0^{ds}0^{ud})
+2\,\alpha_1^{0^{us}} I_1(0^{ds}0^{us})
+2\,\alpha_1^{0^{ds}} I_1(0^{ds}0^{ds})
\nonumber\\
&&\nonumber\\
&+&4\,\alpha_2^{1^{ss}0^{ud}} I_2(0^{ds}0^{ud}1^{ss})
\nonumber\\
&&\nonumber\\
\alpha_2^{1^{uu}1^{ss}}&=&\lambda
+2\,\alpha_1^{1^{ss}} I_4(1^{ss}1^{uu}1^{ss})
+2\,\alpha_1^{0^{ud}} I_4(1^{uu}1^{ss}0^{ud})
+4\, \alpha_1^{0^{us}} I_3(1^{uu}1^{ss}0^{us})
\nonumber\\
&&\nonumber\\
&+&4\,\alpha_2^{1^{ss}0^{ud}} I_6(1^{uu}1^{ss}0^{ud}1^{ss})
\nonumber\\
&&\nonumber\\
\alpha_2^{1^{ss}0^{ud}}&=&\lambda
+4\,\alpha_1^{1^{uu}} I_4(0^{ud}1^{ss}1^{uu})
+2\,\alpha_1^{1^{ss}} I_4(1^{ss}0^{ud}1^{ss})
+\alpha_1^{0^{ud}} I_4(0^{ud}1^{ss}0^{ud})
+2\, \alpha_1^{0^{us}} I_3(1^{ss}0^{ud}0^{us})
\nonumber\\
&&\nonumber\\
&+&2\, \alpha_1^{0^{ds}} I_3(1^{ss}0^{ud}0^{ds})
+2\,\alpha_2^{1^{uu}1^{ss}} I_6(1^{ss}0^{ud}1^{ss}1^{uu})
+2\,\alpha_2^{1^{ss}0^{ud}} I_6(1^{ss}0^{ud}1^{ss}0^{ud})
\nonumber\\
&&\nonumber\\
&+&2\,\alpha_3^{1^{uu}1^{ss}0^{ds}} I_8(1^{ss}0^{ud}1^{ss}0^{ds}1^{uu})
\nonumber\\
&&\nonumber\\
\alpha_3^{1^{uu}1^{ss}0^{ds}}&=&\lambda
+2\, \alpha_1^{1^{ss}} I_9(0^{ds}1^{ss}1^{uu}1^{ss})
+2\, \alpha_1^{0^{ud}} I_9(1^{uu}0^{ds}1^{ss}0^{ud})
+\alpha_1^{0^{us}} (2\, I_9(1^{uu}0^{ds}1^{ss}0^{us})
\nonumber\\
&&\nonumber\\
&+&4\, I_9(1^{uu}1^{ss}0^{ds}0^{us}))
+2\, \alpha_1^{0^{ds}} I_9(0^{ds}1^{ss}1^{uu}0^{ds})
+4\, \alpha_2^{1^{ss}0^{ud}} I_{10}(1^{uu}0^{ds}1^{ss}0^{ud}1^{ss})
\nonumber
\end{eqnarray}

\newpage

Table I. S-wave dibaryon masses. Parameters of model: cutoff
$\Lambda=11.0$, gluon coupling constants $g_0=0.653$ and $g_1=0.292$.
Quark masses $m_{u,d}=410\, MeV$ and $m_s=557\, MeV$.

\vskip2ex

\begin{tabular}{|c|c|c|c|c|c|}
\hline
$S$ & $I$ & Quark content & $J$ & Dibaryon & Mass (MeV) \\
\hline
$0$  & $0$   & $uuuddd$ & $1$       & $NN$               & $1865$ \\
     &       &          & $1$       & $\Delta\Delta$     & $1834$ \\
     & $1$   & $uuuudd$ & $2$       & $N\Delta$, $\Delta\Delta$ & $2020$ \\
     & $2$   & $uuuuud$ & $1$       & $N\Delta$, $\Delta\Delta$ & $1984$ \\
     & $3$   & $uuuuuu$ & $0, \, 2$ & $\Delta\Delta$     & $2379$ \\
\hline
$-1$ & $1/2$ & $uuudds$ & $1$       & $\Delta\Sigma$, $\Delta\Sigma^*$
                                                         & $1936$ \\
     &       &          & $1$       & $N\Sigma$, $N\Sigma^*$ & $1947$ \\
     &       &          & $1$       & $N\Lambda$         & $2024$ \\
     & $3/2$ & $uuuuds$ & $0$       & $\Delta\Sigma^*$   & $1959$ \\
     &       &          & $0$       & $N\Sigma$          & $2080$ \\
     &       &          & $2$       & $\Delta\Sigma$, $\Delta\Sigma^*$
                                                         & $2075$ \\
     &       &          & $2$       & $\Delta\Lambda$    & $2157$ \\
     &       &          & $2$       & $N\Sigma^*$        & $2239$ \\
     & $5/2$ & $uuuuus$ & $1$       & $\Delta\Sigma$, $\Delta\Sigma^*$
                                                         & $2101$ \\
\hline
$-2$ & $0$   & $uuddss$ & $0$       & $\Sigma\Sigma$, $\Sigma^*\Sigma^*$
                                                         & $2118$ \\
     &       &          & $0$       & $\Lambda\Lambda$   & $2173$ \\
     &       &          & $0$       & $N\Xi$             & $2252$ \\
     &       &          & $2$       & $N\Xi^*$           & $2368$ \\
     &       &          & $2$       & $\Sigma\Sigma^*$, $\Sigma^*\Sigma^*$
                                                         & $2411$ \\
     & $1$   & $uuudss$ & $1$       & $\Sigma\Sigma$, $\Sigma\Sigma^*$,
$\Sigma^*\Sigma^*$, $\Lambda\Sigma$, $\Lambda\Sigma^*$   & $2138$ \\
     &       &          & $1$       & $\Delta\Xi$, $\Delta\Xi^*$ & $2292$ \\
     &       &          & $1$       & $N\Xi^*$           & $2336$ \\
     & $2$   & $uuuuss$ & $0$       & $\Sigma\Sigma$, $\Sigma^*\Sigma^*$
                                                         & $2270$ \\
     &       &          & $0$       & $\Delta\Xi^*$      & $2432$ \\
     &       &          & $2$       & $\Delta\Xi$, $\Delta\Xi^*$ & $2432$ \\
     &       &          & $2$       & $\Sigma\Sigma^*$, $\Sigma^*\Sigma^*$
                                                         & $2472$ \\
\hline
$-3$ & $1/2$ & $uudsss$ & $0$       & $\Sigma\Xi$, $\Sigma^*\Xi^*$
                                                         & $2166$ \\
     &       &          & $0$       & $\Lambda\Xi$       & $2243$ \\
     &       &          & $2$       & $\Sigma^*\Xi$, $\Sigma\Xi^*$,
                                    $\Sigma^*\Xi^*$      & $2421$ \\
     &       &          & $2$       & $\Lambda\Xi^*$     & $2481$ \\
     &       &          & $2$       & $N\Omega$          & $2573$ \\
     & $3/2$ & $uuusss$ & $1$       & $\Sigma\Xi$, $\Sigma^*\Xi$,
                          $\Sigma\Xi^*$, $\Sigma^*\Xi^*$ & $2195$ \\
     &       &          & $1$       & $\Delta\Omega$     & $2669$ \\
\hline
$-4$ & $0$   & $udssss$ & $1$       & $\Xi\Xi$, $\Xi\Xi^*$, $\Xi^*\Xi^*$
                                                         & $2428$ \\
     &       &          & $1$       & $\Lambda\Omega$    & $2553$ \\
     & $1$   & $uussss$ & $0$       & $\Xi\Xi$, $\Xi^*\Xi^*$ & $2509$ \\
     &       &          & $0$       & $\Sigma^*\Omega$   & $2706$ \\
     &       &          & $2$       & $\Sigma\Omega$, $\Sigma^*\Omega$
                                                         & $2706$ \\
     &       &          & $2$       & $\Xi\Xi^*$, $\Xi^*\Xi^*$ & $2720$ \\
\hline
$-5$ & $1/2$ & $usssss$ & $1$       & $\Xi\Omega$, $\Xi^*\Omega$ & $2587$ \\
\hline
$-6$ & $0$   & $ssssss$ & $0, \, 2$ & $\Omega\Omega$     & $3232$ \\
\hline
\end{tabular}

\newpage

Table II. S-wave dibaryon masses. Parameters of model: cutoff
$\Lambda=11.0$, gluon coupling constants $g_0=0.647$ and $g_1=0.325$.
Quark masses $m_{u,d}=410\, MeV$ and $m_s=557\, MeV$.

\vskip2ex

\begin{tabular}{|c|c|c|c|c|c|}
\hline
$S$ & $I$ & Quark content & $J$ & Dibaryon & Mass (MeV) \\
\hline
$0$  & $0$   & $uuuddd$ & $1$       & $NN$               & $1848$ \\
     &       &          & $1$       & $\Delta\Delta$     & $1817$ \\
     & $1$   & $uuuudd$ & $2$       & $N\Delta$, $\Delta\Delta$ & $1885$ \\
     & $2$   & $uuuuud$ & $1$       & $N\Delta$, $\Delta\Delta$ & $1941$ \\
     & $3$   & $uuuuuu$ & $0, \, 2$ & $\Delta\Delta$     & $2278$ \\
\hline
$-1$ & $1/2$ & $uuudds$ & $1$       & $\Delta\Sigma$, $\Delta\Sigma^*$
                                                         & $1926$ \\
     &       &          & $1$       & $N\Sigma$, $N\Sigma^*$ & $1938$ \\
     &       &          & $1$       & $N\Lambda$         & $2016$ \\
     & $3/2$ & $uuuuds$ & $0$       & $\Delta\Sigma^*$   & $1940$ \\
     &       &          & $0$       & $N\Sigma$          & $2063$ \\
     &       &          & $2$       & $\Delta\Sigma$, $\Delta\Sigma^*$
                                                         & $2054$ \\
     &       &          & $2$       & $\Delta\Lambda$    & $2144$ \\
     &       &          & $2$       & $N\Sigma^*$        & $2218$ \\
     & $5/2$ & $uuuuus$ & $1$       & $\Delta\Sigma$, $\Delta\Sigma^*$
                                                         & $2057$ \\
\hline
$-2$ & $0$   & $uuddss$ & $0$       & $\Sigma\Sigma$, $\Sigma^*\Sigma^*$
                                                         & $2115$ \\
     &       &          & $0$       & $\Lambda\Lambda$   & $2171$ \\
     &       &          & $0$       & $N\Xi$             & $2242$ \\
     &       &          & $2$       & $N\Xi^*$           & $2360$ \\
     &       &          & $2$       & $\Sigma\Sigma^*$, $\Sigma^*\Sigma^*$
                                                         & $2404$ \\
     & $1$   & $uuudss$ & $1$       & $\Sigma\Sigma$, $\Sigma\Sigma^*$,
$\Sigma^*\Sigma^*$, $\Lambda\Sigma$, $\Lambda\Sigma^*$   & $2129$ \\
     &       &          & $1$       & $\Delta\Xi$, $\Delta\Xi^*$ & $2283$ \\
     &       &          & $1$       & $N\Xi^*$           & $2325$ \\
     & $2$   & $uuuuss$ & $0$       & $\Sigma\Sigma$, $\Sigma^*\Sigma^*$
                                                         & $2241$ \\
     &       &          & $0$       & $\Delta\Xi^*$      & $2397$ \\
     &       &          & $2$       & $\Delta\Xi$, $\Delta\Xi^*$ & $2397$ \\
     &       &          & $2$       & $\Sigma\Sigma^*$, $\Sigma^*\Sigma^*$
                                                         & $2440$ \\
\hline
$-3$ & $1/2$ & $uudsss$ & $0$       & $\Sigma\Xi$, $\Sigma^*\Xi^*$
                                                         & $2156$ \\
     &       &          & $0$       & $\Lambda\Xi$       & $2234$ \\
     &       &          & $2$       & $\Sigma^*\Xi$, $\Sigma\Xi^*$,
                                    $\Sigma^*\Xi^*$      & $2409$ \\
     &       &          & $2$       & $\Lambda\Xi^*$     & $2470$ \\
     &       &          & $2$       & $N\Omega$          & $2565$ \\
     & $3/2$ & $uuusss$ & $1$       & $\Sigma\Xi$, $\Sigma^*\Xi$,
                          $\Sigma\Xi^*$, $\Sigma^*\Xi^*$ & $2170$ \\
     &       &          & $1$       & $\Delta\Omega$     & $2643$ \\
\hline
$-4$ & $0$   & $udssss$ & $1$       & $\Xi\Xi$, $\Xi\Xi^*$, $\Xi^*\Xi^*$
                                                         & $2409$ \\
     &       &          & $1$       & $\Lambda\Omega$    & $2535$ \\
     & $1$   & $uussss$ & $0$       & $\Xi\Xi$, $\Xi^*\Xi^*$ & $2478$ \\
     &       &          & $0$       & $\Sigma^*\Omega$   & $2667$ \\
     &       &          & $2$       & $\Sigma\Omega$, $\Sigma^*\Omega$
                                                         & $2667$ \\
     &       &          & $2$       & $\Xi\Xi^*$, $\Xi^*\Xi^*$ & $2683$ \\
\hline
$-5$ & $1/2$ & $usssss$ & $1$       & $\Xi\Omega$, $\Xi^*\Omega$ & $2530$ \\
\hline
$-6$ & $0$   & $ssssss$ & $0, \, 2$ & $\Omega\Omega$     & $3093$ \\
\hline
\end{tabular}

\newpage

\noindent
Table III. Vertex functions and Ghew-Mandelstam coefficients.

\vskip3ex

\begin{tabular}{|c|c|c|c|c|}
\hline
$i$ & $G_i^2(s_{kl})$ & $\alpha_i$ & $\beta_i$ & $\delta_i$ \\
\hline
& & & & \\
$0^+$ & $\frac{4g}{3}-\frac{8gm_{kl}^2}{(3s_{kl})}$
& $\frac{1}{2}$ & $-\frac{1}{2}\frac{(m_k-m_l)^2}{(m_k+m_l)^2}$ & $0$ \\
& & & & \\
$1^+$ & $\frac{2g}{3}$ & $\frac{1}{3}$
& $\frac{4m_k m_l}{3(m_k+m_l)^2}-\frac{1}{6}$
& $-\frac{1}{6}\frac{(m_k-m_l)^2}{(m_k+m_l)^2}$ \\
& & & & \\
\hline
\end{tabular}

\newpage

{\bf \Large References.}

\vskip5ex

\noindent
1. R.L. Jaffe, Phys. Rev. Lett. {\bf 38}, 195 (1977).

\noindent
2. F. Wang, J.L. Ping, H.R. Pang and T. Goldman, Mod. Phys. Lett. A{\bf 18},
356 (2003).

\noindent
3. G.H. Wu, J.L. Ping, F. Wang and T. Goldman, Nucl. Phys. A{\bf 673},
279 (2000).

\noindent
4. M. Oka, K. Shimizu and K. Yazaki, Phys. Lett. B{\bf 130}, 365 (1983).

\noindent
5. P.J.G. Mulders, A.T. Aerts and J.J. Swarts, Phys. Rev. Lett. {\bf 40},
1543 (1978).

\noindent
6. A. Faessler, F. Fernandez, G. Lubeck et al., Nucl. Phys. A{\bf 402},
555 (1983).

\noindent
7. I.T. Obukhovsky and A.M. Kusainov, Phys. Lett. B{\bf 238}, 142 (1990).

\noindent
8. E.M. Henley and C.A. Miller, Phys. Lett. B{\bf 251}, 453 (1991).

\noindent
9. T. Kamae and T. Fujita, Phys. Rev. Lett. {\bf 38}, 471 (1977).

\noindent
10. K. Yazaki, Prog. Theor. Phys. Suppl. {\bf 91}, 146 (1987).

\noindent
11. F. Wang, G.H. Wu, L.J. Teng and T. Goldman, Phys. Rev. Lett. {\bf 69},
2901 (1992).

\noindent
12. T. Goldman, K. Maltman, G.J. Stephenson Jr, J.-L. Ping and F. Wang,

Mod. Phys. Lett. A{\bf 13}, 59 (1998).

\noindent
13. Z.Y. Zhang, Y.W. Yu, X.Q. Yuan et al., Nucl. Phys. A{\bf 670}, 178
(2000).

\noindent
14. V.B. Kopeliovich, Nucl. Phys. A{\bf 639}, 75 (1998).

\noindent
15. P. LaFrance and E.L. Lomon, Phys. Rev. D{\bf34}, 1341 (1986).

\noindent
16. Y.W. Yu, Z.Y. Zhang and X.Q. Yuan, Communi. Theor. Phys. {\bf 31},
1 (1999).

\noindent
17. Y.W. Yu, Z.Y. Zhang and X.Q. Yuan, High Energy Phys. and Nucl. Phys.
{\bf 23}, 859 (1999).

\noindent
18. I.J.R. Aitchison, J. Phys. G{\bf 3}, 121 (1977).

\noindent
19. J.J. Brehm, Ann. Phys. (N.Y.) {\bf 108}, 454 (1977).

\noindent
20. I.J.R. Aitchison and J.J. Brehm, Phys. Rev. D{\bf 17}, 3072 (1978).

\noindent
21. I.J.R. Aitchison and J.J. Brehm,
Phys. Rev. D{\bf 20}, 1119, 1131 (1979).

\noindent
22. J.J. Brehm, Phys. Rev. D{\bf 21}, 718 (1980).

\noindent
23. S.M. Gerasyuta and E.E. Matskevich, Phys. Rev. D{\bf 76}, 116004 (2007).

\noindent
24. S.M. Gerasyuta and E.E. Matskevich, Int. J. Mod. Phys. E{\bf 17},
585 (2008).

\noindent
25. S.M. Gerasyuta and E.E. Matskevich, Yad. Fiz. {\bf 70}, 1995 (2007).

\noindent
26. A. De Rujula, H. Georgi and S.L. Glashow, Phys. Rev. D{\bf 12}, 147
(1975).

\noindent
27. G.'t Hooft, Nucl. Phys. B{\bf 72}, 461 (1974).

\noindent
28. G. Veneziano, Nucl. Phys. B{\bf 117}, 519 (1976).

\noindent
29. E. Witten, Nucl. Phys. B{\bf 160}, 57 (1979).

\noindent
30. O.A. Yakubovsky, Sov. J. Nucl. Phys. {\bf 5}, 1312 (1967).

\noindent
31. S.P. Merkuriev and L.D. Faddeev, Quantum Scattering Theory for System
of Few Particles

(Nauka, Moscow 1985) p. 398.

\noindent
32. Y. Nambu and G. Jona-Lasinio, Phys. Rev. {\bf 122}, 345 (1961);
{\bf 124}, 246 (1961).

\noindent
33. T. Appelqvist and J.D. Bjorken, Phys. Rev. D{\bf 4}, 3726 (1971).

\noindent
34. C.C. Chiang, C.B. Chiu, E.C.G. Sudarshan and X. Tata,
Phys. Rev. D{\bf 25}, 1136 (1982).

\noindent
35. V.V. Anisovich, S.M. Gerasyuta and A.V. Sarantsev,
Int. J. Mod. Phys. A{\bf 6}, 625 (1991).

\noindent
36. G. Chew, S. Mandelstam, Phys. Rev. {\bf 119}, 467 (1960).

\noindent
37. V.V. Anisovich and A.A. Anselm, Usp. Fiz. Nauk {\bf 88}, 287 (1966).

\noindent
38. S.M. Gerasyuta, Z. Phys. C{\bf 60}, 683 (1993).

\noindent
39. Q.B. Li, P.N. Shen, Z.Y. Zhang, Y.W. Yu, Nucl. Phys. A{\bf 683}, 487
(2001).

\noindent
40. M. Bashkanov et al., Prog. Part. Nucl. Phys. {\bf 61}, 304 (2008).

\noindent
41. H. Pang, J. Ping, F. Wang, T. Goldman and E. Zhao, Phys. Rev. C{\bf 69},
065207 (2004).

\end{document}